\documentclass{pasj01}
\Received{$\langle$reception date$\rangle$}
\Accepted{$\langle$acception date$\rangle$}
\Published{$\langle$publication date$\rangle$}
\usepackage[colorlinks=true,urlcolor=blue,linkcolor=blue,citecolor=blue]{hyperref}

\usepackage{natbib}
\usepackage{aas_macros}

\usepackage{url}
\usepackage{ulem}

\begin{document}

\title{21 cm Forest Constraints on Primordial Black Holes}
\author{Pablo Villanueva-Domingo\altaffilmark{1}}
\author{Kiyotomo Ichiki \altaffilmark{2,3}}
\altaffiltext{1}{Instituto de F\'isica Corpuscular (IFIC),
  CSIC-Universitat de Val\`encia,\\
Apartado de Correos 22085,  E-46071, Spain}
\altaffiltext{2}{Graduate School of Science, Division of Particle and
Astrophysical Science, Nagoya University, Chikusa-Ku, Nagoya, 464-8602, Japan}
\altaffiltext{3}{Kobayashi-Maskawa Institute for the Origin of Particles and the
Universe, Nagoya University, Chikusa-Ku, Nagoya, 464-8602, Japan}
\email{pablo.villanueva.domingo@gmail.com, ichiki@a.phys.nagoya-u.ac.jp}


\maketitle

\begin{abstract}
Primordial black holes (PBHs) as part of the Dark Matter (DM) would modify the evolution of large-scale structures and the thermal history of the universe. Future 21 cm forest observations, sensitive to small scales and the thermal state of the Inter Galactic Medium (IGM), could probe the existence of such PBHs. In this article, we show that the shot noise isocurvature mode on small scales induced by the presence of PBHs can enhance the amount of low mass halos, or minihalos, and thus, the number of 21 cm absorption lines. However, if the mass of PBHs is as large as $M_{\rm PBH}\gtrsim 10 \, M_\odot$, with an abundant enough fraction of PBHs as DM, $f_{\rm PBH}$, the IGM heating due to accretion onto the PBHs counteracts the enhancement due to the isocurvature mode, reducing the number of absorption lines instead. The concurrence of both effects imprints distinctive signatures in the number of absorbers, allowing to bound the abundance of PBHs. We compute the prospects for constraining PBHs with future 21 cm forest observations, finding achievable competitive upper limits on the abundance as low as $f_{\rm PBH} \sim 10^{-3}$ at $M_{\rm PBH}= 100 \, M_\odot$, or even lower at larger masses, in unexplored regions of the parameter space by current probes. The impact of astrophysical X-ray sources on the IGM temperature is also studied, which could potentially weaken the bounds.
\end{abstract}

\section{Introduction}

After the LIGO collaboration detected gravitational waves from the merging event of intermediate-mass black holes \citep{2016PhRvL.116x1103A}, primordial black holes (PBHs) have attracted much attention again as the possible origin of those merger events \citep{2016PhRvL.117f1101S}. PBHs are interesting objects in cosmology because they can constitute part of the Dark Matter (DM) in the universe \citep{2017PDU....15..142C,2016PhRvD..94h3504C}. Furthermore, their abundance can be employed to investigate the amplitude of primordial density fluctuations on very small scales \citep{2019JCAP...10..031K} responsible for the PBH formation. They could also act as seeds to form super massive black holes, whose large masses are hardly explained from standard accretion mechanisms \citep{Carr:2018rid}. Many astronomical observations have placed stringent constraints on their abundance over extended ranges of their mass from different probes, including gravitational lensing, gamma-ray background, and cosmic microwave background (CMB) (see, e.g., \cite{2020arXiv200710722G, 2020arXiv200212778C, Villanueva-Domingo:2021spv} for recent reviews regarding the current constraints on PBHs). 

Assuming that the primordial fluctuations are originally Gaussian distributed, PBHs are expected to be intially randomly distributed on small scales, with no significant clustering
\citep{2018PhRvL.121h1304A, 2018PhRvD..98l3533D}. If the spatial distribution of PBHs in the universe follows the Poisson statistics, given that they behave as discrete objects, the existence of PBHs will enhance cosmological density fluctuations on small scales \citep{2003ApJ...594L..71A}. Therefore, observations of small-scale structures can be used to constrain the PBH abundance in the universe, as done, e.g., via the Ly$\alpha$ forest \citep{2019PhRvL.123g1102M}. A promising probe could be provided by observing the cosmological 21 cm line, one of the most propitious tools to study the intergalactic medium (IGM) from the Cosmic Dawn to the Epoch of Reionization \citep{Pritchard:2011xb, Furlanetto:2006jb}. Gong \& Kitajima investigated the 21 cm line emission signal of the neutral hydrogen in low mass halos, known as minihalos, from the density fluctuations enhanced by PBHs in the Reionization epoch. 
They found that future 21 cm line surveys such as Square Kilometre Array (SKA) can constrain the fraction to DM density to be less than $10^{-3}$ for $10$ $ M_\odot$ mass PBHs \citep{Gong:2017sie, Gong:2018sos}. On the other hand, \cite{Mena:2019nhm} further investigated the 21 cm signal in PBHs scenarios, emitted or absorbed in the IGM as well as in minihalos. They found that the enhancement of the signal present in \cite{Gong:2018sos} due to the shot noise power spectrum could be significantly diminished when taking into account the heating of the IGM, either from astrophysical sources or from accretion onto PBHs. The reason is that the minimum collapsed mass, which can be estimated as the Jeans mass, grows with the temperature. Hence, a hot IGM would increase the minimum mass of minihalos, suppressing their number and counteracting the enhancement of the power spectrum at small scales.

This paper investigates the prospect of constraining the mass and abundance of PBHs using 21 cm line observations. Unlike the references mentioned above, which accounted for the signal emitted (or absorbed) relative to the CMB from minihalos or in the IGM, we study the absorption lines that constitute the so-called 21 cm forest.  The 21 cm forest is the absorption line system in the spectra of bright radio sources caused by the intervening neutral hydrogen atoms, in an analogous way to the Ly$\alpha$ forest \citep{2002ApJ...577...22C, 2002ApJ...579....1F, Xu_2009, 2011MNRAS.410.2025X, 2012MNRAS.425.2988M, Semelin:2015wca}. Unlike observations of the 21 cm temperature from the IGM (either its global average or the power spectrum) with the CMB as the backlight, the 21 cm forest is observed against a bright radio source, allowing easy removal of foregrounds, which is the main problem of the IGM signal. Furthermore, it allows to reach smaller scales since it only depends upon the frequency resolution of the observed spectra. Moreover, the 21 cm forest can probe earlier times and smaller scales than the Ly$\alpha$ forest can do owing to its smaller optical depth. Prospects for detecting the 21 cm forest with LOFAR and SKA have been examined in the literature \citep{2013MNRAS.428.1755C, Ciardi:2015lia}. Several works have made use of the 21 cm forest in order to obtain constraints on decaying DM \citep{2012arXiv1205.1204V}, neutrino masses, running spectral index, or warm DM masses \citep{2014PhRvD..90h3003S}, ultra-light DM particles \citep{2020PhRvD.101d3516S, Kawasaki:2020tbo} or axion DM \citep{2020PhRvD.102b3522S}. We compute the prospects for observing the 21 cm forest within a PBH scenario. We incorporate the enhancement at small scales from the Poisson shot noise power spectrum, as done in \cite{Gong:2017sie, Gong:2018sos}. Moreover, we consider the effect of gas accretion onto PBHs, as properly accounted in \cite{Mena:2019nhm}, that was neglected in the former works. 

This paper is organized as follows. In section \ref{sec:forest}, we briefly review the 21 cm forest methodology. In section \ref{sec:PBHs}, we discuss the effects of the PBHs on the matter power spectrum and the heating of the IGM. We present our results in section \ref{sec:results}, whereas section \ref{sec:conclusions} is devoted to the summary and discussion of the most relevant conclusions.

\section{21 cm forest}
\label{sec:forest}
During the Cosmic Dawn and the Epoch of Reionization, the neutral hydrogen atoms that create 21 cm absorption lines resided mostly in the IGM or the DM halos. In \cite{2002ApJ...579....1F}, the authors showed that distinctive absorption lines should be created due to neutral hydrogen atoms in the minihalos. In contrast, the IGM forms continuous absorptions with a small optical depth observed as an overall declination in the amplitude of the radio spectrum. Minihalos are those DM halos not massive enough to host star formation within. They typically present virial temperatures below $10^4$ K, which corresponds to the atomic cooling threshold to produce effective star formation activity in Population (Pop) II stars \citep{Barkana:2000fd}\footnote{Note that early metal-free Pop III stars with molecular cooling could have lower virial temperatures, down to $\sim 10^3$ K.}. Therefore, one can expect the abundant neutral hydrogen atoms existing in minihalos to create deep absorption lines. In this work, we only consider the absorption lines due to neutral hydrogen in minihalos, whose number density in the universe depends on the amplitude of cosmological density perturbations and hence on the number and mass of the PBHs. In this section, we show how to calculate the number of absorption lines due to minihalos. We mostly base our methodology on the seminal work by \cite{2002ApJ...579....1F}.

\subsection{Minihalos}

\begin{figure*}[th!]
\begin{center}
	\includegraphics[width=0.49\linewidth, bb=0.00 0.00 420.52 316.01]{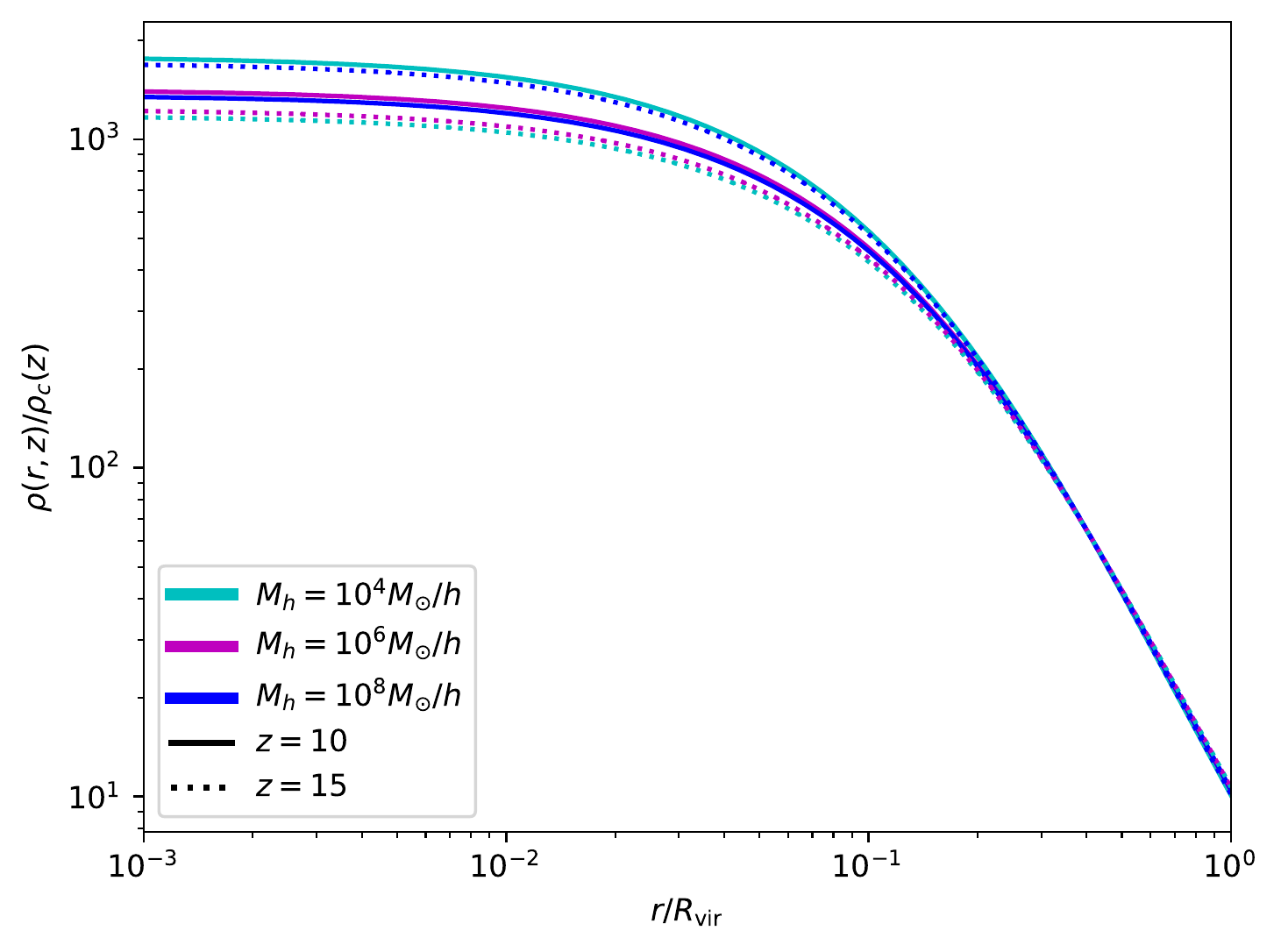}
	\includegraphics[width=0.49\linewidth, bb=0.00 0.00 420.52 316.01]{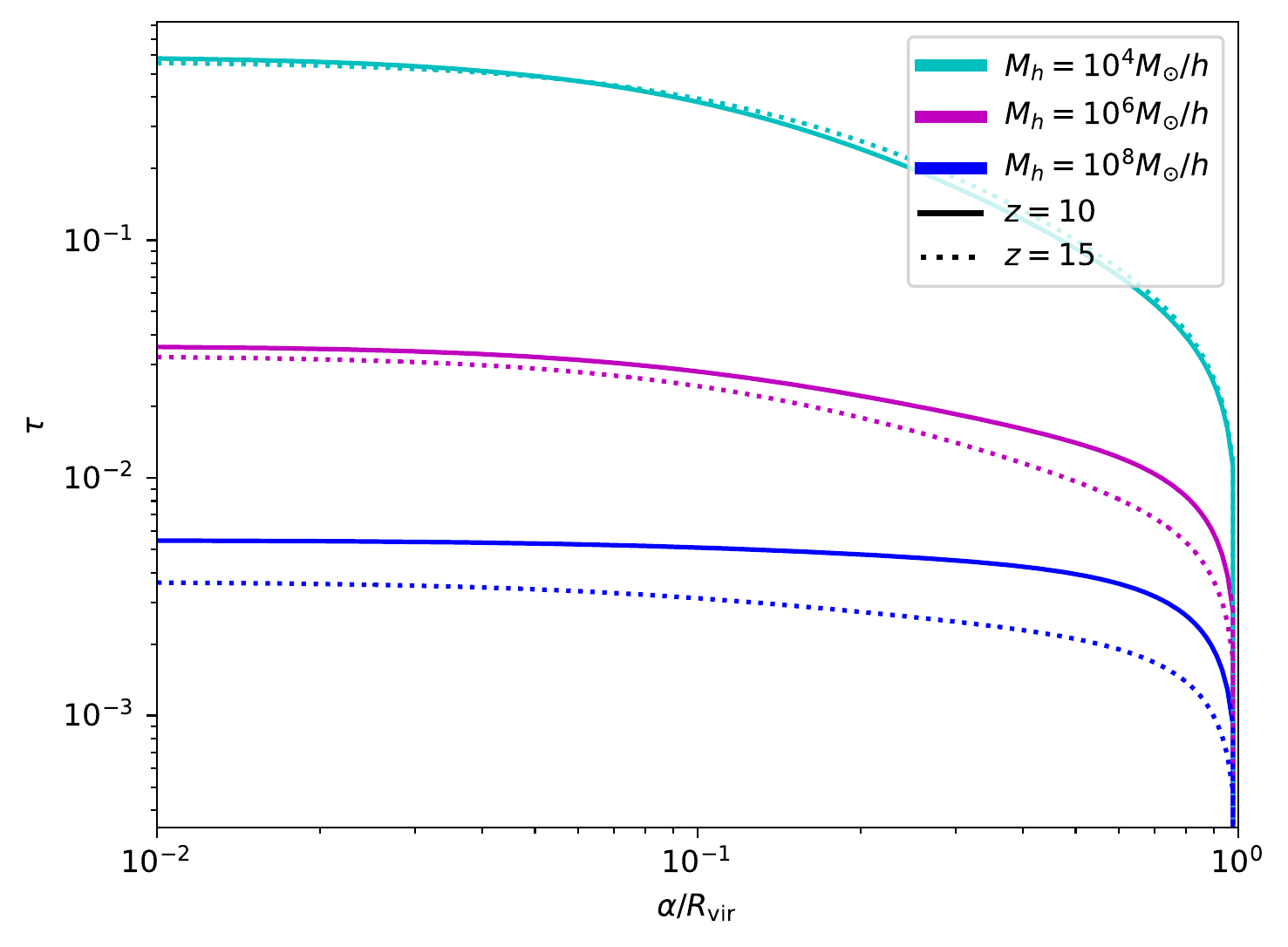}
	\caption{\textit{Left:} density profile of the gas as a function of radius. \textit{Right:} 21 cm optical depth as a function of the impact parameter. Both panels show results for redshifts 10 and 15, as well as for 3 different halo masses.}
	\label{fig:taurho}
	\end{center}
\end{figure*}

The virial temperature $T_{\rm vir}$ of the DM halo with mass $M_{\rm h}$ is given by \citep{Barkana:2000fd}
\begin{eqnarray}
  T_{{\rm vir}} &=&1.98\times 10^{4} \,{\rm K}\, \bigg(\frac{\mu}{0.6}\bigg)\bigg(\frac{M_h}{10^{8}h^{-1}M_{\odot}}\bigg)^{2/3}\nonumber \\
  &&\times \bigg[\frac{\Omega_{m}}{\Omega_{m}^{z}}\frac{\Delta_{c}}{18\pi^{2}}\bigg]^{1/3}\bigg(\frac{1+z}{10}\bigg)~,
\end{eqnarray}
where $\mu$ is the mean molecular weight, $h$ is the dimensionless Hubble parameter, $\Omega_m$ is the cosmological matter density parameter, and $\Delta_{c}=18\pi ^{2}+82d-39d^{2}$ is the overdensity of virialized halos collapsing at redshift ${\it z}$, with $d=\Omega_{m}^{z}-1$ and
$\Omega_{m}^{z} = \Omega_{m}(1+z)^{3}/(\Omega_{m}(1+z)^{3}+\Omega_{\Lambda})$ \citep{Bryan:1997dn}. 
The virial radius $R_{{\rm vir}}$ is defined through its relation with the mass of the halo $M_h$, $M_h = \frac{4\pi}{3} \, \Delta_c(z) \, \rho_c(z) \, R_{\rm vir}^3$, with $\rho_c(z)$ the critical density at redshift $z$, being given therefore by
\begin{eqnarray}
  R_{{\rm vir}}&=&0.784 \,\,{\rm kpc}\, \bigg(\frac{M_h}{10^{8}h^{-1}M_{\odot}}\bigg)^{1/3}\bigg[\frac{\Omega_{m}}{\Omega_{m}^{z}}
  \frac{\Delta_{c}} {18\pi^{2}}\bigg]^{-1/3} \nonumber \\
  &&\times \bigg(\frac{1+{\it z}}{10}\bigg)^{-1}h^{-1}~.
\label{eq:virial_radius}
\end{eqnarray}

We assume the NFW profile for the DM distribution in a minihalo \citep{Navarro:1996gj}:
\begin{equation}
\rho_{\rm NFW}(r) = \frac{y^3}{3 F(y)} \frac{\Delta_c \rho_c(z)}{\frac{r}{r_s}(1+\frac{r}{r_s})^2}~,
\end{equation}
where $r_s$ is the scaling radius, given by the concentration $y = R_{vir}/r_s$, and $F(x)=\ln(1+x)-x/(1+x)$. Therefore, its integral up to a radius $r$ provides the total mass $M(r)$ as
\begin{equation}
M(r)=4\pi \int_0^r dr' r'^2 \rho_{\rm NFW}(r) = M_{h} \frac{F(r/r_s)}{F(y)}~.
\end{equation}
For the concentration parameter, we make use of a numerical fit to match N-body simulations as a function of the mass and the redshift. Specifically, we employ the prescription provided by \cite{Ishiyama:2020vao}, based on the semianalytical model proposed in \cite{Diemer:2018vmz}. We make use of its implementation in the Python package {\tt Colossus} \citep{2018ApJS..239...35D}. The form adopted applies for a broad range of halo masses and for the redshifts considered here, and mostly agrees with other parameterizations of the concentration (see, e.g.,  \cite{Maccio:2008pcd, Ragagnin:2018enf, Child:2018skq})\footnote{Note that some previous 21 cm forest calculations \citep{2014PhRvD..90h3003S, 2020PhRvD.101d3516S, 2020PhRvD.102b3522S,  Kawasaki:2020tbo} made use of the parameterization from \cite{2001MNRAS.321..559B}, which shows a stronger dependence with redshift, and increases the optical depth at low impact parameters and halo masses. This thus provides an enhancement of the number of absorbers at large optical depths, see Sec. \ref{sec:numabs}.}.

In order to compute the gas density profile, we consider a simple scenario with an isothermal distribution, fixing the temperature to $T_k=T_{{\rm vir}}$. Furthermore, we only consider the DM as the source for the gravitational potential. Imposing hydrostatic equilibrium, we can find an equation for the gas density $\rho_g$ by equating the pressure gradient with the gravitational force. Defining the pressure as $P=T_k \rho_g/(\mu m_p)$, we have
\begin{equation}
\frac{dP}{dr} = \frac{T_k}{\mu m_p}\frac{d\rho_g}{dr} = -\rho_g \frac{G M(r)}{r^2}~.
\end{equation}
The equation above can be easily solved, getting \citep{1998ApJ...497..555M}
\begin{equation}
   \rho_{ g}(r)= \rho_{ g,0} \exp{\left[-\frac{\mu m_{{\rm p}}}{2k_{{\rm B}}T_{{\rm vir}}}(V_{{\rm esc}}^{2}(0)-V_{{\rm esc}}^{2}(r)) \right]},
\label{eq:gas_profile1}
\end{equation}
where $\rho_{g,0}$ is the value at the center and the escape velocity $V_{\rm esc}$ is given by
\begin{equation}
  V_{{\rm esc}}^{2}(r) = 2\int_{r}^{\infty}\frac{GM(r')}{r'^{2}}dr'=2V_{c}^{2}\frac{F(yx)+yx/(1+yx)}{xF(y)}~,
\label{eq:esc_velocity}
\end{equation}
where $x\equiv r/R_{{\rm vir}}$ and $V_{c}$ is the circular velocity, which reads \citep{Barkana:2000fd}
\begin{eqnarray}
  V_{c}^{2} =  \frac{GM_h}{R_{{\rm vir}}} &=&23.4 \, \mbox{km/s} \bigg(\frac{M_h}{10^{8}h^{-1}M_{\odot}}\bigg)^{1/3}\nonumber \\
  &&\times \bigg[\frac{\Omega_{m}}{\Omega_{m}^{z}}\frac{\Delta_{c}}{18\pi^{2}}\bigg]^{1/6}
 \bigg(\frac{1+z}{10}\bigg)^{1/2}~.
\label{eq:cir_velocity}
\end{eqnarray}
Note that $V_{c}^{2}=2k_B T_{\rm vir}/(\mu m_p)$. The center of the density profile, $\rho_{g,0}$, can be obtained by imposing that the total gas mass within the virial radius is $M_{gas} = (\Omega_b/\Omega_m) M_h$,
\begin{equation}
  \rho_{g,0}(z) =
  \frac{(\Delta_c/3)y^{3}e^{A}}{\int_{0}^{y}(1+t)^{A/t}t^{2}dt} \left(\frac{\Omega_b}{\Omega_m}\right)\rho_{c}(z)~,
\end{equation}
where $A=V_{{\rm esc}}^{2}(0)/V_{c}^{2}=2y/F(y)$ \footnote{Note that in some previous works \citep{2014PhRvD..90h3003S, 2020PhRvD.102b3522S}, the quantity $A$ presents a factor of 3 rather than 2, which may come from an additional factor of $3/2$ in the definition of the virial temperature.}. 
We depict the density profile of the HI gas in the left panel of Fig.~(\ref{fig:taurho}), where one can see that the profile only shows a weak dependence with the mass of the halo and the redshift for the ranges considered. Note that at $z=15$, the density increases with the halo mass, while at $z=10$, this behavior is reversed, presenting larger densities the halos with lower masses. This non-trivial phenomenon arises from the mass-redshift dependence of the concentration function from \cite{Ishiyama:2020vao}, which slightly increases (decreases) with larger masses at $z=15$ ($z=10$). Nonetheless, all the cases match at far enough distances from the center, where the background limit is recovered, independently on the mass and redshift.

\subsection{21 cm optical depth}

The relative number of triplet $n_1$ and singlet $n_0$ states of the hyperfine structure of the hydrogen is quantified by the spin temperature $T_S$, implicitly defined as $n_1/n_0=3\, {\rm exp}(-h\nu_{21}/T_S)$, with the factor of 3 accounting for the internal degrees of freedom. Three main processes drive its evolution: i) absorption and emission of 21 cm photons with respect to the CMB, ii) collisions with hydrogen atoms, electrons, or protons, and iii) resonant scattering of Ly$\alpha$ photons (the so-called Wouthuysen-Field effect). Each of these processes may produce spin-flip transitions, exciting or de-exciting the hydrogen atom, and their balance leads to the expression \citep{Pritchard:2011xb}
\begin{equation}
T_S^{-1}=\frac{T^{-1}_{\gamma}+x_cT^{-1}_k+x_{\alpha}T^{-1}_c}{1+x_c+x_{\alpha}}~,
\label{eq:spintemp}
\end{equation}
with $T_{\gamma}=2.73 \, {\rm K}(1+z)$ the CMB temperature, $T_c \simeq T_k$ is the effective color temperature of the radiation field around the Ly$\alpha$ line, and $x_c$ and $x_\alpha$ are the coupling coefficients for collisions and Ly$\alpha$ scattering, respectively. It is not clear whether the Ly$\alpha$ flux coming from the IGM would be able to reach the inner zones of the minihalo, or it would be shelf-shielded against that radiation. Following previous works \citep{2014PhRvD..90h3003S, 2020PhRvD.101d3516S, Kawasaki:2020tbo}, we make the simplifying approximation of neglecting the Ly$\alpha$ coupling effect inside the minihalo, assuming that its spin temperature is only given by the first two processes commented above. As already commented, inside the minihalos, we set $T_k = T_{\rm vir}$. We thus expect that the spin temperature gets coupled to the kinetic one in the inner parts of the minihalo, where densities are large and collisional coupling dominates, while approaching the CMB one in the outer zones. Neglecting the Ly$\alpha$ coupling channel can be seen as a conservative assumption, given that, if the IGM is not heated, its kinetic temperature would be cold, and $T_S$ would also be lower in the outer regions. It would increase the optical depths, being thus easier to detect, which would imply stronger bounds. On the other hand, if the IGM is heated by astrophysical or accretion processes, the difference with our approach would not be so significant. Moreover, the largest optical depths arise from the inner parts of the minihalo (corresponding to small impact parameters, as explained below). At these central regions, collisional coupling is expected to dominate, due to the large overdensities, being the Ly$\alpha$ coupling only relevant at the outer regions. Hence, the largest values of the optical depth would be mainly given by collisional coupling. Given that minihalos are neutral, hydrogen-hydrogen collisions constitute the dominant channel. For the collisional coupling coefficient, we make use of the fitting formulae from \cite{2006ApJ...637L...1K}.

The absorption of 21 cm photons is determined by the opacity at a radius $r$ from the center of the minihalo, $\kappa(\nu,r)$ (e.g., \cite{2019cosm.book.....M}),
\begin{equation}
\kappa(\nu,r) = \frac{h\nu_{21}}{4\pi}B_{01}\frac{1}{4}n_{\rm HI}(r)\frac{h\nu_{21}}{T_S}\phi(\nu)~.
\end{equation}
Here, $B_{01}$ is the absorption Einstein coefficient, $\nu_{21}=1420$ MHz is the frequency of the hyperfine transition, $n_{\rm HI}$ is the neutral hydrogen number density and $\phi(\nu)={\rm exp}(-v^{2}(\nu)/b^2)/(\sqrt{\pi} \nu_{21} b/c)$ is the line profile, where we consider the Doppler broadening effect with $v(\nu)=c(\nu-\nu_{21})/\nu_{21}$ and the velocity dispersion $b=\sqrt{2k_B T_k/m_p}$. The factor $1/4$ accounts for the fraction of neutral atoms in the singlet state of the hyperfine structure of the hydrogen. Given an impact parameter $\alpha$ of a ray that penetrates a minihalo, the 21 cm optical depth is computed integrating from $-R_{{\rm max}}$ to $R_{{\rm max}}$, where $R_{{\rm max}}=\sqrt{R_{\rm vir}^2 - \alpha^2}$,
 \begin{eqnarray}
   \tau(\nu,M,\alpha)&=& \int_{-R_{{\rm max}}(\alpha)}^{R_{{\rm max}}(\alpha)} dR \; \kappa(\nu,R) \nonumber \\
   &\simeq& \int_{-R_{{\rm max}}(\alpha)}^{R_{{\rm max}}(\alpha)} dR \frac{h\nu_{21}}{4\pi}B_{01}\frac{1}{4}n_{\rm HI}(r)\frac{h\nu}{T_S}\phi(\nu)	\\
   &=&\frac{3h_{{\rm p}}c^{3}A_{10}}{32\pi k_{{\rm B}}\nu_{21}^{2}}
   \int_{-R_{{\rm max}}(\alpha)}^{R_{{\rm max}}(\alpha)}dR\frac{n_{{\rm HI}}(r)}{T_{{\rm S}}(r)\sqrt{\pi}b}
   e^{-v^{2}(\nu)/b^2}, \nonumber
 \label{eq:optical1}
 \end{eqnarray}
where $r^2 = R^2 + \alpha^2$. The optical depths $\tau(\nu,M,\alpha)$ for halos with some masses in redshift $z=10$ and $z=15$ as a function of the impact parameter $\alpha$ are depicted in the right panel of Fig.~(\ref{fig:taurho}). One can clearly see that low mass halos with small impact parameters lead to larger optical depths. 

\subsection{Number of absorbers}
\label{sec:numabs}

For a given mass scale $M$ and an optical depth $\tau$, we can compute the maximum impact parameter $r_{\tau}(M,\tau)$. The cumulative abundance of absorption features per redshift interval is then written as
\begin{equation}
\label{eq:mass_integration}
  \frac{dN(>\tau)}{dz}=\frac{dr}{dz}\int_{M_{{\rm min}}}^{M_{{\rm max}}}dM\frac{dn}{dM}\pi r^{2}_{\tau}(M,\tau), \label{eq:abundance1}
\end{equation}
where $dr/dz$, $r_{\tau}$ and the halo mass function $dn/dM$ (defined in the next section) are in comoving units. Note that the factor $\pi r^{2}_{\tau}$ acts as an effective cross section\footnote{Writing the cross section in physical units gives the factor $(1+z)^2$ of \cite{2002ApJ...579....1F}.}. Thus, Eq. (\ref{eq:abundance1}) accounts for the number of absorbers (per redshift interval) which provides absorption troughs with optical depths larger than $\tau$. The lower limit of the integral is set by the lowest mass of collapsed objects, which is given by the Jeans mass, $M_{\rm J}$, the mass within a region of diameter\footnote{Note that there are different conventions to define the Jeans Mass, which differ in an order $\mathcal{O}(1)$ factor.}  $\lambda_J = \sqrt{\left(5\pi k_{\rm B}T_{\rm IGM}\right)/\left(3G\bar{\rho}m_{\rm p}\mu\right)}$,
\begin{eqnarray}
M_{\rm J}(T_k,z) &=& \frac{4 \pi}{3}  \bar{\rho} \left( \frac{\lambda_{\rm J}}{2} \right)^3 \nonumber \\
%&\simeq& 2.2 \times 10^9 M_\odot  \left(\frac{T_k}{10^4~{\rm K}}\right)^{3/2} \, \left(\frac{1.22}{\mu} \right)^{3/2} \, \left(\frac{10}{1+z}\right)^{3/2} ~,
%&\simeq& 2.2 \times 10^3 M_\odot  \left(\frac{T_k}{1~{\rm K}}\right)^{3/2} \, \left(\frac{1.22}{\mu} \right)^{3/2} \, \left(\frac{10}{1+z}\right)^{3/2} ~.
&\simeq& 1.3 \times 10^3 M_\odot h^{-1}  \left(\frac{T_k}{1~{\rm K}}\right)^{3/2} \,  \left(\frac{10}{1+z}\right)^{3/2} ~,
\label{eq:jeans}
\end{eqnarray}
assuming the molecular weight as $\mu=1.22$. On the other hand, the upper limit is fixed by the virial mass corresponding to the virial temperature as high as $\sim 10^4$ K, which indicates the minimum mass to host star formation via atomic cooling:
\begin{eqnarray}
M_{\rm vir}^{\rm min}(T_{\rm vir},z)&\simeq& 10^8 M_{\odot}h^{-1} \left( \frac{T_{\rm vir}}{1.98 \times 10^4 K}\right)^{3/2} \nonumber \\
&&\times \bigg[\frac{\Omega_{m}}{\Omega_{m}^{z}}\frac{\Delta_{c}}{18\pi^{2}}\bigg]^{-1/2}\bigg(\frac{1+z}{10}\bigg)^{-3/2}.
\end{eqnarray}
Halos with larger masses are expected to produce stars within, becoming ionized, and thus not contributing to the 21 cm forest.

It is worth to note that one could apply a version of Eq. (\ref{eq:abundance1}) for other features of the 21 cm forest besides the optical depth, such as the intrinsic equivalent width \citep{2002ApJ...579....1F}. For simplicity, here we focus only on $\tau$ as the characteristic quantity of the 21 cm forest.

\subsection{Detectability of the 21 cm forest}

How likely is it to observe the 21 cm forest in the future? Such detections may need relatively bright radio background fluxes to distinguish the absorption troughs in the spectra. In this section we estimate the required sources where 21 cm forest observations are feasible.

The signal-to-noise ratio SNR can be written as the brightness temperature of the observed spectra $T$ divided by the root-mean-square noise $T_{\rm rms}$. Such noise temperature is given by the system temperature $T_{\rm{sys}}$ (which includes the one from electronics and the sky temperature) divided by the standard deviation of the number of data points, which can be estimated as $\sqrt{\Delta \nu t_{\rm int}}$, where $\Delta \nu$ is the frequency resolution and $t_{\mathrm{int}}$ is the observation time. Thus, $T_{\rm rms}=T_{\rm{sys}}/\sqrt{\Delta \nu t_{\rm int}}$ and ${\rm SNR} = (T/T_{\rm{sys}})\sqrt{\Delta \nu t_{\rm int}}$. On the other hand, one can relate the flux $S$ with the brightness temperature within the Rayleigh-Jeans approximation as $S=2k_BT\Omega/\lambda^2$, with $\lambda$ the wavelength and $\Omega$ the subtended solid angle. Both single antennas and radiointerferometers fulfill the condition $\Omega \simeq \lambda^2/A_{\rm{eff}}$, where $A_{\rm{eff}}$ is the effective collecting area. Therefore, from the previous relations, the minimum detectable flux density of a radio source with a specified signal-to-noise ratio SNR can be written as
\citep{2011MNRAS.410.2025X}
\begin{equation}
\Delta S=\frac{2 k_B T_{\rm{sys}}}{A_{\rm{eff}}\sqrt{\Delta \nu t_{{\rm int}}}}\, {\rm SNR}.
\end{equation}
Hence, in order to distinguish between the emitted flux $S$ and the one damped by the 21 cm forest, $S e^{-\tau}$, the experiment has to be sensitive to $\Delta S = S(1-e^{-\tau})\simeq S\tau$. Equating with the formula above, one finds the minimum brightness of a radio source required to observe the 21 cm forest with absorption features characterized by a target 21 cm optical depth  $\tau$ as \citep{2002ApJ...579....1F, Furlanetto:2006dt}
\begin{eqnarray}
S_{\rm min}=10.4 \, \rm{mJy} && \, \left(\frac{0.01}{\tau}\right)\left(\frac{{\rm SNR}}{5}\right)\left(\frac{1 \, \rm{kHz}}{\Delta \nu}\right)^{1/2}\\
 &&\times\left(\frac{10^6 {\rm m^{2}}}{A_{\rm{eff}}}\right)\left(\frac{T_{\rm{sys}}}{200 \rm K}\right)\left(\frac{100~\rm{hr}}{t_{{\rm int}}}\right)^{1/2},
\label{eq:Smin}
\end{eqnarray}
where the fiducial values correspond to the SKA specifications observing at $z\sim 10$, which imply a minimum required flux of order ${\cal O}(1 \sim 10)$ mJy. For comparison, an equivalent to the powerful local radio source Cygnus A located at $z\sim 10$ would emit a flux of $\lesssim 20$ mJy \citep{Furlanetto:2006jb, 2012MNRAS.425.2988M}.

The main uncertainty regarding the 21 cm forest is whether there are such bright radio sources at high redshifts. There has been significant progress recently finding new radio bright sources at $z\gtrsim 6$ such as quasars \citep{ban2015, ban2018, Belladitta:2020zbp, Banados:2021imw}. Only four out of the $\sim 200$ known quasars at $z > 6$ are known to be radio-loud (i.e., where radio emission is the dominant component in their spectra), being the farthest one recently discovered at $z=6.82$ \citep{Banados:2021imw}. However, around 10 \% of all quasars could be radio-loud, nearly independently on redshift \citep{Ivezic:2002gh, ban2015}. The brightest radio source known so far is a radio-loud quasar found at $z = 5.84$, with the flux $8\sim 100$ mJy from $3$ GHz to $230$ MHz, which can be bright enough for the 21 cm forest studies if its local dense gas environment is confirmed by the follow-up surveys \citep{ban2018}. The estimates based on extrapolations of the observed radio luminosity functions to the higher redshift indicate that there could be as many as $\sim 10^{4}-10^{5}$ radio quasars with sufficient brightness at $z \sim 10$ \citep{Haiman_2004,Xu_2009}.

Gamma-ray bursts (GRBs) may also present a radio afterglow where the 21 cm forest could be potentially identified \citep{2005ApJ...619..684I, Ciardi:2015lia}. There are several known GRBs at high redshift (at least 6 at $z > 6 $), such as the ones at $z=6.7$ \citep{Greiner:2008fb}, $z=8.2$ \citep{2009Natur.461.1254T} and $z=9.4$ \citep{2011ApJ...736....7C}. Pop III stars in metal-free environments may lead to the formation of GRBs much more energetic than the ordinary ones, being able to generate much brighter low-frequency radio afterglows, exceeding tens of mJy \citep{2011ApJ...731..127T, 2011A&A...533A..32D}. The claimed detection of the 21 cm global signal by EDGES \citep{2018Natur.555...67B}, if  verified, would imply that Pop III stars should exist at $z\gtrsim 17$ \citep{2018MNRAS.480L..43M}, as well as Pop III GRBs at $z\gtrsim 10$. Hence, all these findings and estimates present good prospects for observing the 21 cm forest in the future.

\section{Effects of Primordial Black Holes}
\label{sec:PBHs}

In this section, two of the most relevant effects of solar mass PBHs to cosmology are discussed, namely the enhancement of fluctuations due to shot noise and the heating of the IGM due to accretion. Throughout this work, we assume a monochromatic distribution for the PBH mass at $M_{\rm PBH}$ and write the fraction of PBHs that comprise DM as $f_{\rm PBH}=\Omega_{\rm PBH}/\Omega_{\rm DM}$, with $\Omega_{\rm PBH}$ and $\Omega_{\rm DM}$ the current energy density of PBHs and DM respectively relative to the critical density $\rho_{c,0}$.

\subsection{Shot noise}

\begin{figure*}[ht!]
 \begin{center}
 	\includegraphics[width=0.49\textwidth]{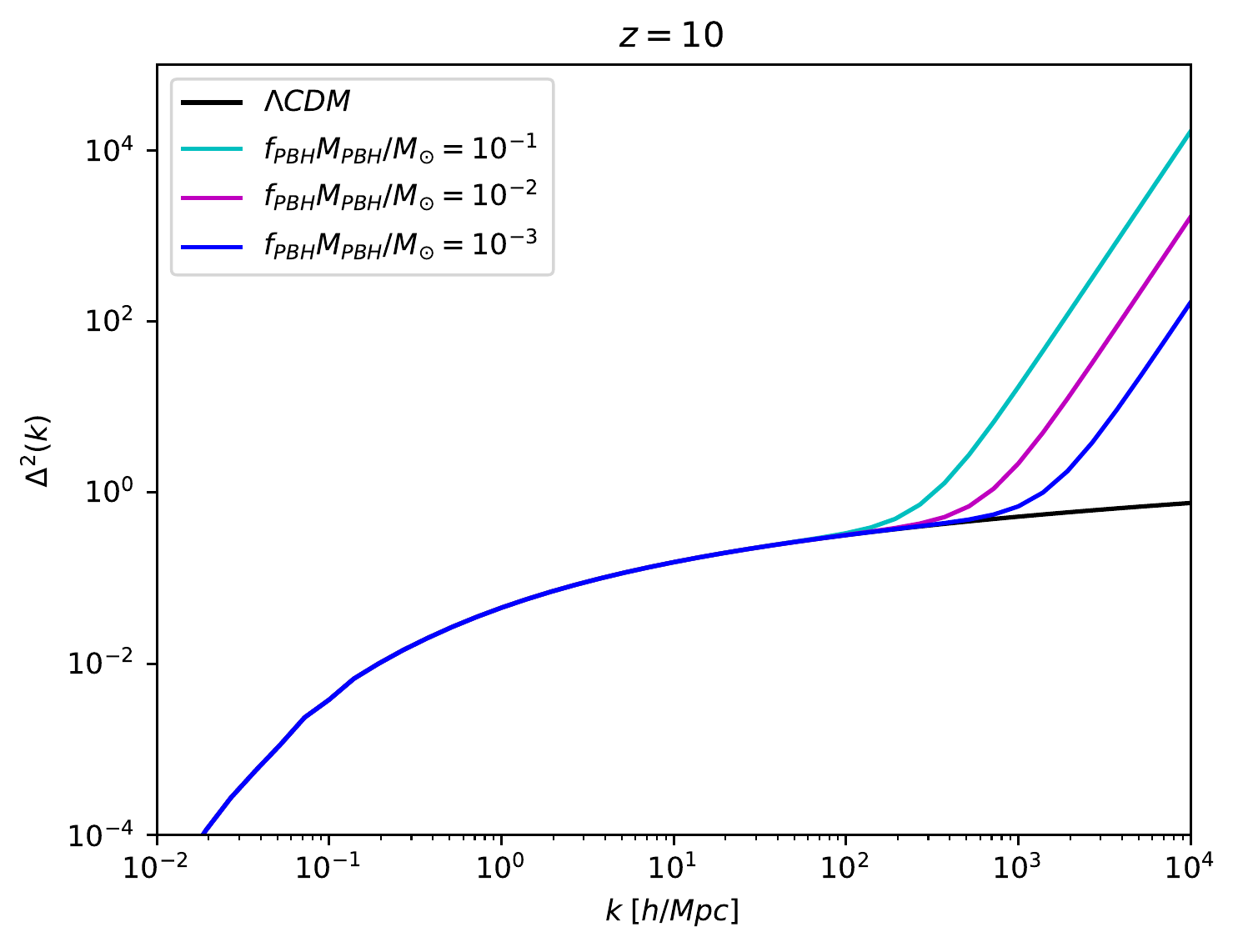}
 	\includegraphics[width=0.49\textwidth]{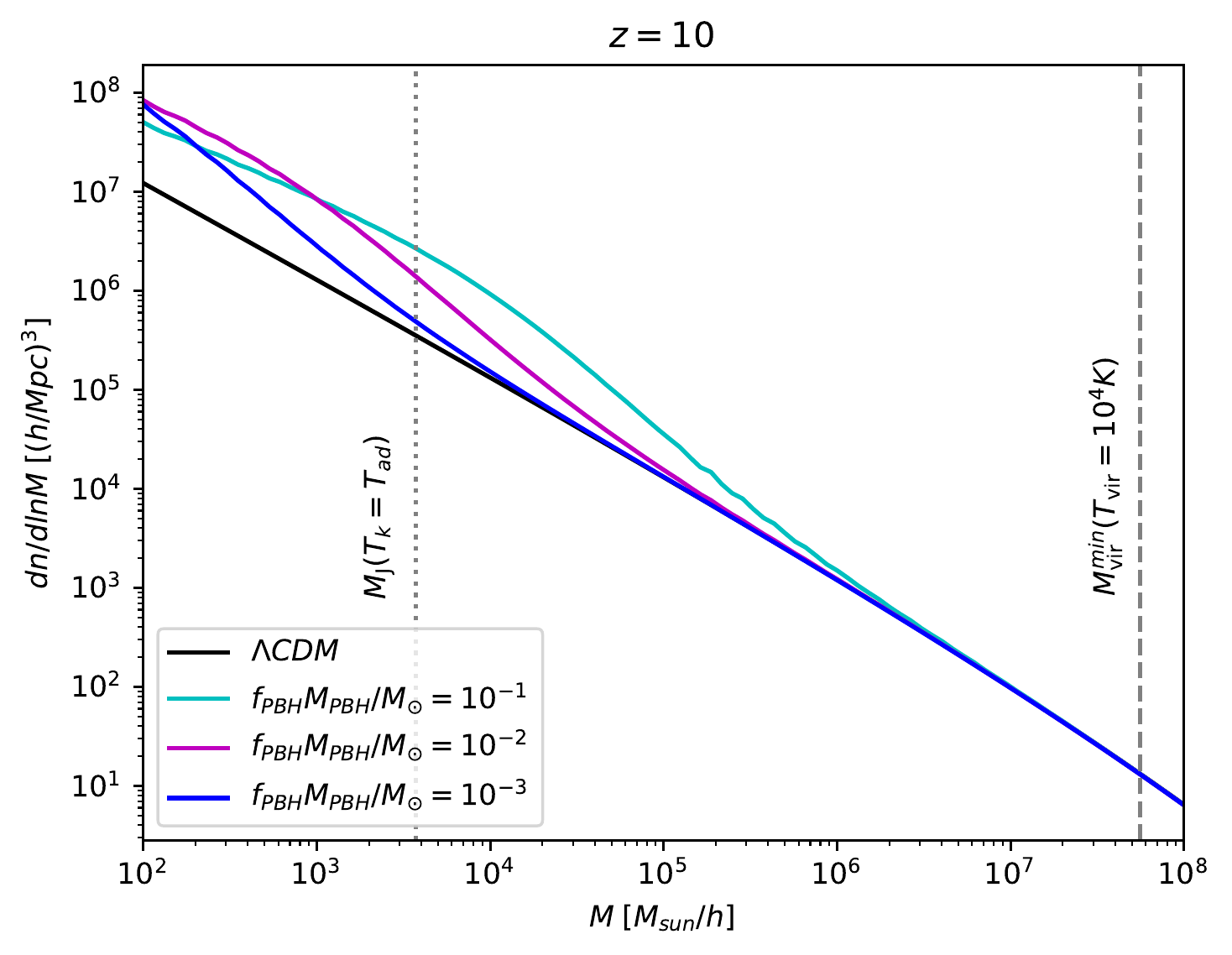}
 	\caption{Matter power spectrum (left) and halo mass function (right) for several values of $f_{\rm PBH}M_{\rm PBH}$ at $z=10$. The vertical lines in the number of halos panel correspond to the Jeans mass evaluated at the adiabatic temperature (\textit{i.e.}, without considering IGM heating) and the mass relative to a virial temperature of $10^4$ K, respectively. Note that when including the accretion effects outlined in Sec. \ref{sec:acc}, the IGM is heated up and the Jeans mass becomes larger.}
 	\label{fig:PSdndM}
 	\end{center}
 \end{figure*}

Since PBHs can be regarded as discrete objects, if they are initially Poisson distributed, they should present a shot noise power spectrum independent on the wavenumber $k$ (i.e., white noise) and inversely proportional to their number density. The Poisson shot noise contribution in the matter power spectrum given by the discrete distribution of PBHs is
\begin{equation}
P_{\rm Poisson} = \frac{f_{\rm PBH}^2}{n_{\rm PBH}}~,
\end{equation}
where the number density of PBHs $n_{\rm PBH}$ can be written as
\begin{equation}
n_{\rm PBH} = \frac{f_{\rm PBH} \Omega_{\rm DM} \rho_{c,0}}{M_{\rm PBH}}.
\end{equation}
As was noted in \cite{Afshordi:2003zb} (see also \cite{2006PhRvD..73h3504C, Inman:2019wvr}), the density perturbations generated by PBHs correspond to isocurvature modes. The total matter power spectrum shall thus be given by a combination of isocurvature and isentropic fluctuations. The transfer function for isocurvature modes can be approximated by (see, \textit{e.g.}, \cite{Peacock:1999ye})
\begin{equation}
\mathcal{T}_{\rm iso} = \frac{3}{2}(1+z_{eq})~,
\end{equation}
for $k>k_{\rm eq}$, and $\mathcal{T}_{\rm iso}\simeq 0$ otherwise, with $z_{\rm eq}$ the redshift of matter-radiation equality, and $k_{\rm eq} = H(z_{\rm eq})/(1+z_{\rm eq})$. Hence, the contribution of isocurvature fluctuations only affects scales smaller than those corresponding to the
matter-radiation equality era. Therefore, the total contribution to the linear power spectrum at redshift $z=0$ is
\begin{eqnarray}
P_{\rm iso} = \mathcal{T}^2_{\rm iso} P_{\rm Poisson} &=&  \frac{9}{4}(1+z_{\rm eq})^2\frac{f_{\rm PBH}M_{\rm PBH}}{\Omega_{\rm DM} \rho_c} \\ 
&\simeq & 2.5 \times 10^{-2} f_{\rm PBH} \left(\frac{M_{\rm PBH}}{30 \, M_\odot}\right) \textrm{Mpc}^3,
\end{eqnarray}
for $k>k_{\rm{eq}}$, being 0 otherwise. Note that the above expression only depends on the joint product of the PBH fraction and the mass, being this contribution degenerate on these two parameters. For a generic redshift $z$, the power spectrum is written as $P_{\rm iso}(k,z)=D^2(z)\mathcal{T}_{\rm iso}^2(k)P_{\rm Poisson}(k)$,  where $D(z)$ is the growth factor of linear perturbations. The dimensionless matter power spectrum, defined as $\Delta^2=k^3 P(k)/(2\pi^2)$ including the isocurvature and adiabatic modes, is shown in the left panel of Fig.~(\ref{fig:PSdndM}), for several PBH parameters and for the standard $\Lambda$CDM case.

The number density of halos per unit mass is given by the halo mass function \citep{Press:1973iz},
\begin{equation}
\label{eq:dndm}
\frac{dn}{dM_{\rm h}} = \frac{\rho_{\rm m}}{M_{\rm h}} \, \frac{d \ln \sigma^{-1}}{d M_{\rm h}} \, f(\sigma) ~,
\end{equation}
where $f(\sigma)$ represents the fraction of mass that has collapsed to form halos per unit interval in $\ln \sigma^{-1}$, with $\sigma$ being the root-mean-square of the matter density fluctuation smoothed with a real-space top hat filter over the virial radius, 
\begin{equation}
\sigma^2(M_h) = \frac{1}{2\pi^2} \int dk \; k^2 P(k) |W(k,M_h)|^2~,
\end{equation}
where $W(k; M_{\rm h})$ is the Fourier transform of the real-space top hat filter. Here we use the Sheth-Tormen halo mass function~\citep{Sheth:1999mn, Sheth:1999su},
\begin{equation}
f(\sigma) = A \, \sqrt{\frac{2 \, a}{\pi}} \, \left[ 1 + \left(\frac{\sigma^2}{a \, \delta_c(z)^2} \right)^p\right]\frac{\delta_c(z)}{\sigma} \, e^{-\frac{a \delta_c(z)^2}{2 \sigma^2}} ~,
\end{equation}
with the fitting coefficients $A = 0.3222$, $a = 0.707$, and $p = 0.3$, and the critical collapse density $\delta_c(z) = 1.686/D(z)$. The halo mass function for several PBH scenarios, together with the standard $\Lambda$CDM model, is shown in the right panel of Fig.~(\ref{fig:PSdndM}). One can see that, for products as large as $f_{\rm PBH}M_{\rm PBH} \sim 10^{-1}$, the isocurvature modes dominates the power on small scales $k\gtrsim 10^2$ $h$/Mpc and thus affects the number of DM halos with small masses $M\lesssim 10^{6}M_\odot$/h.

\subsection{PBH Accretion in the IGM}
\label{sec:acc}

In this section we examinate the impact of accretion by  PBHs on the IGM. Matter infalling onto a PBH would accelerate, and by bremsstrahlung and other processes, would emit energetic radiation, escaping the local vicinity of the PBH and being deposited in the IGM. We follow the procedure from \cite{Mena:2019nhm} for implementing the effects of accretion in the thermal history of the universe, which mostly relies on the previous works by \cite{Ali-Haimoud:2016mbv} and \cite{Poulin:2017bwe}. The rate of energy injected into the medium per unit volume from the accretion of matter onto PBHs is expressed as \citep{Ali-Haimoud:2016mbv, Poulin:2017bwe}
\begin{equation}\label{eq:enginj}
\left(\frac{dE}{dV\,dt}\right)_{\rm inj} = L_{\rm acc} \, n_{\rm PBH} = L_{\rm acc} \, \frac{f_{\rm PBH} \, \rho_{\rm DM}}{M_{\rm PBH}} ~,
\end{equation}
where $L_{\rm acc}$ is the luminosity of radiation emitted from matter accreting onto a PBH of mass $M_{\rm PBH}$ and $\rho_{\rm DM}$ is the density of DM. The accretion luminosity can be generally expressed in terms of the product of the accretion rate $\dot{M}_{\rm PBH}$ and the so-called radiative efficiency $\epsilon$, dependent on $\dot{M}_{\rm PBH}$, as, 
\begin{equation}\label{eq:lacc}
L_{\rm acc} = \epsilon(\dot{M}_{\rm PBH}) \, \dot{M}_{\rm PBH} ~.
\end{equation}
A value of $\epsilon$ close to unity indicates that most of the energy accreted by the PBH is radiated, while lower efficiencies imply that most of the energy remains in the local vicinity of the PBH. Both the accretion rate and the radiative efficiency are complicated functions of the PBH properties and the evolution of the medium. However, various approximations have been developed that can be used to estimate each of these functions.

Firstly, the accretion rate can be easily estimated for spherical accretion, following the seminal computation by \cite{1952MNRAS.112..195B}, as
\begin{equation}
\label{eq:BHrate}
\dot{M}_{\rm PBH} = 4 \pi \, \lambda \, \rho_\infty \, \frac{(G M_{\rm PBH})^2}{v_{\rm eff}^3} ~,
\end{equation}
where $\rho_\infty$ is the density far away from the point source, $G$ is the gravitational constant, and $\lambda$ is the dimensionless accretion rate, a parameter quantifying non-gravitational effects \citep{bondi1944mechanism}, which in spherical accretion takes values $\sim0.1-1$, depending on the equation of state \citep{Ali-Haimoud:2016mbv}. The effective velocity $v_{\rm eff}$ is a combination of the speed of sound far away from the point source $c_{s,\infty}$, and the relative velocity between the PBH and the medium $v_{\rm rel}$ averaged with a Gaussian distribution. We employ the prescription outlined in \cite{Mena:2019nhm} for computing $v_{\rm eff}$; see that article for more details. Although strictly valid only for spherical accretion, the above formula can also apply to more realistic disk accretion, provided that $\lambda$ is reduced by roughly two orders of magnitude, in order to account for viscosity effects and the outflows of matter \citep{Poulin:2017bwe}. Hence, we consider disk accretion by setting $\lambda=0.01$.

On the other hand, the efficiency function $\epsilon$ depends upon the specific accretion scenario. There are several accretion mechanisms depending on the accretion rate, geometry, temperature and properties of the medium. If the radiative cooling is effective and a thin accretion disk forms, the radiative efficiency could be relatively large, with $\epsilon \simeq 0.1$, as in the classic model by \cite{Shakura:1972te}. Here, we assume the so-called  Advective-Dominated-Accretion-Flow (ADAF) model \citep{Ichimaru:1977uf, Rees:1982pe, Narayan:1994xi, Yuan:2014gma}, where the radiative cooling is inefficient, and the dynamics of the accreting gas is controlled by advective currents. In this case, most of the energy remains in the accretion disk, heating it up and leading to the formation of a thicker disk. The ADAF accretion model provide much lower values of the efficiency compared to the \cite{Shakura:1972te} scenario, and is thus more conservative. For the efficiency function, we adopt the fitting formula proposed in \cite{Xie:2012rs},
\begin{equation}
\label{eq:adaffit}
\epsilon = \epsilon_0 \, \left(\frac{10 \, \dot{M}_{\rm PBH}}{L_{\rm Edd}} \right)^a ~,
\end{equation}
where $L_{\rm Edd} = 4 \pi \, G M_{\rm PBH} \, m_p /\sigma_{\rm T}\simeq 1.28 \times 10^{38} (M_{\rm PBH}/M_\odot)$ ergs/s is the Eddington luminosity, and the functional form and coefficients are indicated in Tab. (\ref{table:adaf}), assuming the viscous parameter to be $\alpha = 0.1$ and the parameter that dictates the fraction of dissipation resulting in heating the electrons directly to be $\delta = 0.1$ (see \cite{Xie:2012rs} for more details). It is worth to note that, for the masses and redshifts of interest, the accretion rates always fulfill $\dot{M}_{\rm PBH} / L_{\rm Edd} < 10^{-4}$, and thus, the efficiency belongs to the regime corresponding to the first row of Tab. (\ref{table:adaf}), with a power-law index of $a=0.59$. This corresponds to a specific scenario within ADAF accretion known as electron ADAF (eADAF), where both ions and electrons are radiatively inefficient (see \cite{Yuan:2014gma} for further information).

\begin{table}
	\begin{center}
		\begin{tabular}{|c|c|c|}
			\hline
			$10 \, \dot{M}_{\rm PBH} / L_{\rm Edd}$ & $\epsilon_0$ & $a$ \\ 
			\hline\hline
			$\lesssim 9.4 \times 10^{-5}$ & 0.12 & 0.59 \\
			\hline
			$9.4\times 10^{-5} - 5 \times 10^{-3}$ & 0.026 & 0.27 \\
			\hline
			$5\times 10^{-3} - 6.6 \times 10^{-3}$ & 0.50 & 4.53\\
			\hline
		\end{tabular}
	\end{center}
	\caption{Fit coefficients characterizing the radiative efficiency of ADAF accretion, described by Eq.~(\ref{eq:adaffit}), in terms of the accretion rate. These values are taken from \cite{Xie:2012rs}, for the ADAF electron heating parameter $\delta = 0.1$ and for the viscous parameter $\alpha = 0.1$.}
	\label{table:adaf}
\end{table}

To compute the total heating rate in the IGM produced by accretion, one has to note that energy is deposited also in other channels such as ionization or ${\rm Ly}\alpha$ excitations. Given the injected energy from Eq.~(\ref{eq:enginj}), we can compute the rate of energy deposited into the medium as heating as
\begin{equation}
\label{eq:dep}
\left(\frac{dE_{\rm heat}}{dV dt}\right)_{{\rm dep}} = f_{\rm heat}(z) \, \left(\frac{dE}{dV dt}\right)_{\rm inj} ~,
\end{equation}
where $f_{\rm heat}(z)$ represents the energy deposition fraction in the medium as heat, which is computed by using the following expression:
\begin{equation}\label{eq:energydeposit}
f_{\rm heat}(z) = \frac{H(z)\int \frac{d\ln(1+z^\prime)}{H(z^\prime)} \int d\omega \, T_{\rm heat}(z,z^\prime,\omega) \, L_{\rm acc}(z^\prime, \omega)}{\int d\omega \, L_{\rm acc}(z, \omega)} ~,
\end{equation}
where the $T_{\rm heat}(z,z^\prime,\omega)$ is the transfer function taken from ~\cite{Slatyer:2015kla}. Following \cite{Poulin:2017bwe}, the spectrum of the luminosity is fitted from the ADAF models of \cite{Yuan:2014gma}, adopting a simple parameterization given by 
\begin{equation}\label{eq:L_spec}
L(\omega) \propto \Theta(\omega - \omega_{\rm min}) \, \omega^{c} \, \exp(-\omega / \omega_s) ~,
\end{equation}
where $\omega_{\rm min} \equiv (10 \, M_\odot / M_{\rm PBH})^{1/2}$ eV, $\omega_s \sim \mathcal{O}(m_e)$ (taken explicitly to be $\omega_s = 200$~keV), and $c = -1$ (with reasonable values of $c \in [-1.3, -0.7]$). 

Finally, the evolution of the kinetic temperature of the gas $T_k$ is computed via
\begin{equation}
\label{eq:dTkdzacc}
\nonumber \frac{dT_k}{dz} = \frac{2 \, T_k}{3 \, n_b} \frac{dn_b}{dz} - \frac{T_k}{1 + x_e} \frac{dx_e}{dz} + \frac{2}{3 \, k_B \, (1 + x_e)} \frac{dt}{dz} \sum_j \mathcal{Q}_j ~,
\label{eq:heating}
\end{equation}
where the first term of the right hand side accounts for adiabatic cooling due to the expansion of the universe, the second one includes a contribution from the change in ionizations, and the last term accounts for the heating/cooling processes, namely, Compton cooling,  astrophysical X-ray heating and X-ray heating from PBH accretion, which can be written as $\mathcal{Q}_{\rm PBH}=\left(\frac{dE_{\rm heat}}{dV dt}\right)_{{\rm dep}}/n_b$ from Eq.~(\ref{eq:dep}).

Finally, it is worth to examinate the scaling of the heating rate with the PBH parameters. From Eq. (\ref{eq:BHrate}), one has $\dot{M}_{\rm PBH}\propto M^2_{\rm PBH}$. Noting that $L_{\rm Edd} \propto M_{\rm PBH}$, the fit of Eq.~(\ref{eq:adaffit}) for the efficiency scales as $\epsilon \propto (\dot{M}_{\rm PBH}/M_{\rm PBH})^a \propto M_{\rm PBH}^{a}$. Hence, from Eq.~(\ref{eq:enginj}), $L_{\rm acc} \propto M_{\rm PBH}^{a+2}$, and finally, the rate of energy injected from
Eq.~(\ref{eq:lacc}) goes as $\left(\frac{dE_{\rm heat}}{dV dt}\right)_{{\rm inj}} \propto f_{\rm PBH} M_{\rm PBH}^{a+1}$. Taking into account that the accretion rates considered correspond to the fit for the efficiency given by the first row of Tab. (\ref{table:adaf}), the injection of energy depends upon the combination $\sim f_{\rm PBH} M_{\rm PBH}^{1.59}$. Given that the energy deposition fraction depends weakly with the mass, the heating term $\mathcal{Q}_{\rm PBH}$ and the kinetic temperature should show the same scaling, as is explicitly checked in Sec. \ref{sec:temperatureIGM}.

\section{Results and Discussion}
\label{sec:results}

In this section, the main results of our computations are discussed in detail. The number of absorbers in PBH scenarios, taking into account shot noise and accretion effect, is calculated to estimate the possible limits on the PBH abundance that future observations could achieve. The codes to perform these computations have been released to be publicly available \citep{pablo_villanueva_domingo_2021_4707447}\footnote{\url{https://github.com/PabloVD/21cmForest_PBH}}.

\subsection{Thermal evolution of the IGM}
\label{sec:temperatureIGM}

 \begin{figure}[t]
 \begin{center}
 	\includegraphics[width=1.0\linewidth]{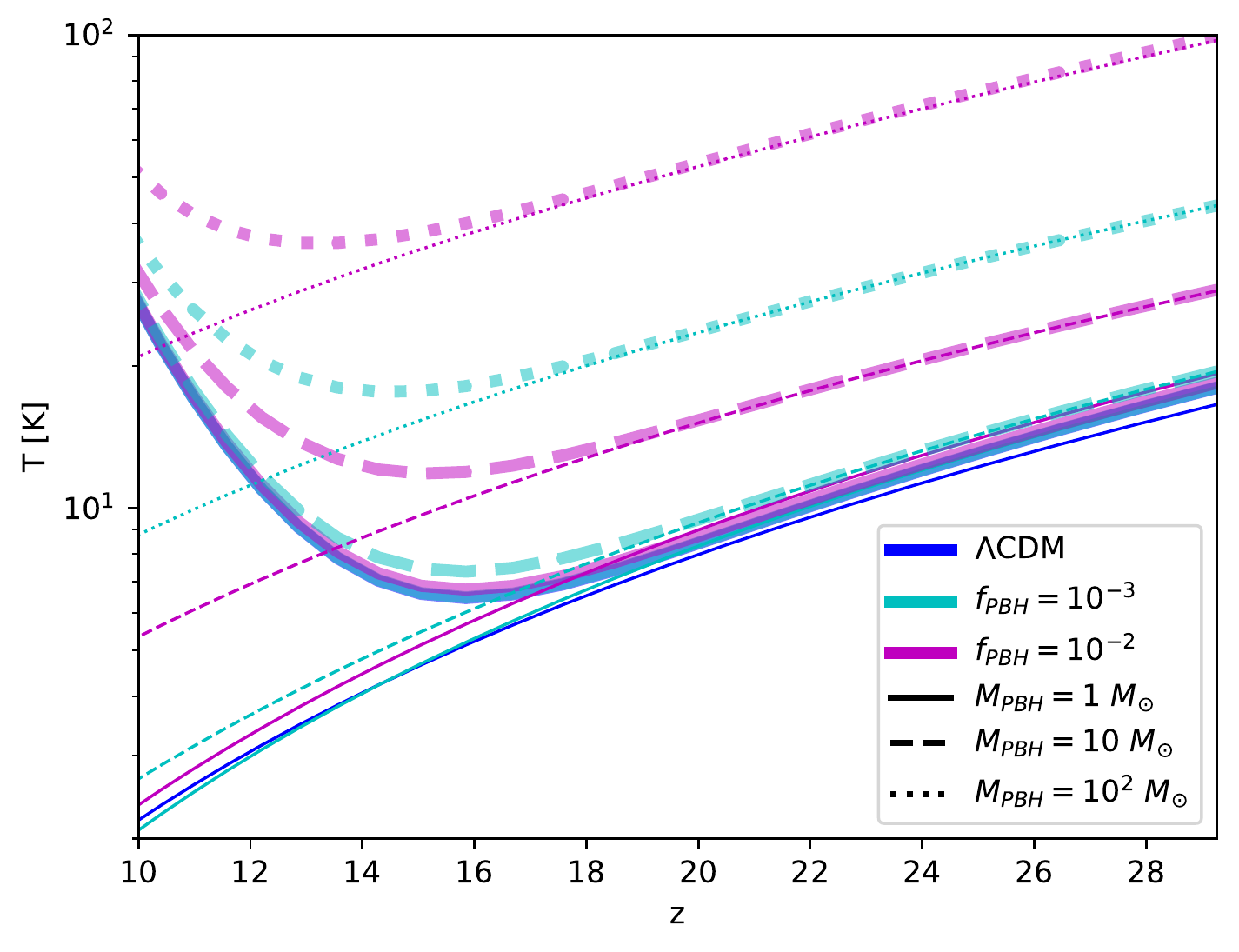}
 	\caption{IGM temperature evolution with redshift for a CDM scenario and for several PBH models. Thin lines account for a scenario without astrophysical X-ray heating, presenting only heating by PBH accretion, while thick lines correspond to a model with $L_X/{\rm SFR}= 10^{40}$ erg s$^{-1}$. }
 	\label{fig:temperature}
 	\end{center}
 \end{figure}
 
\begin{figure}[t]
 \begin{center}
 	\includegraphics[width=1.0\linewidth]{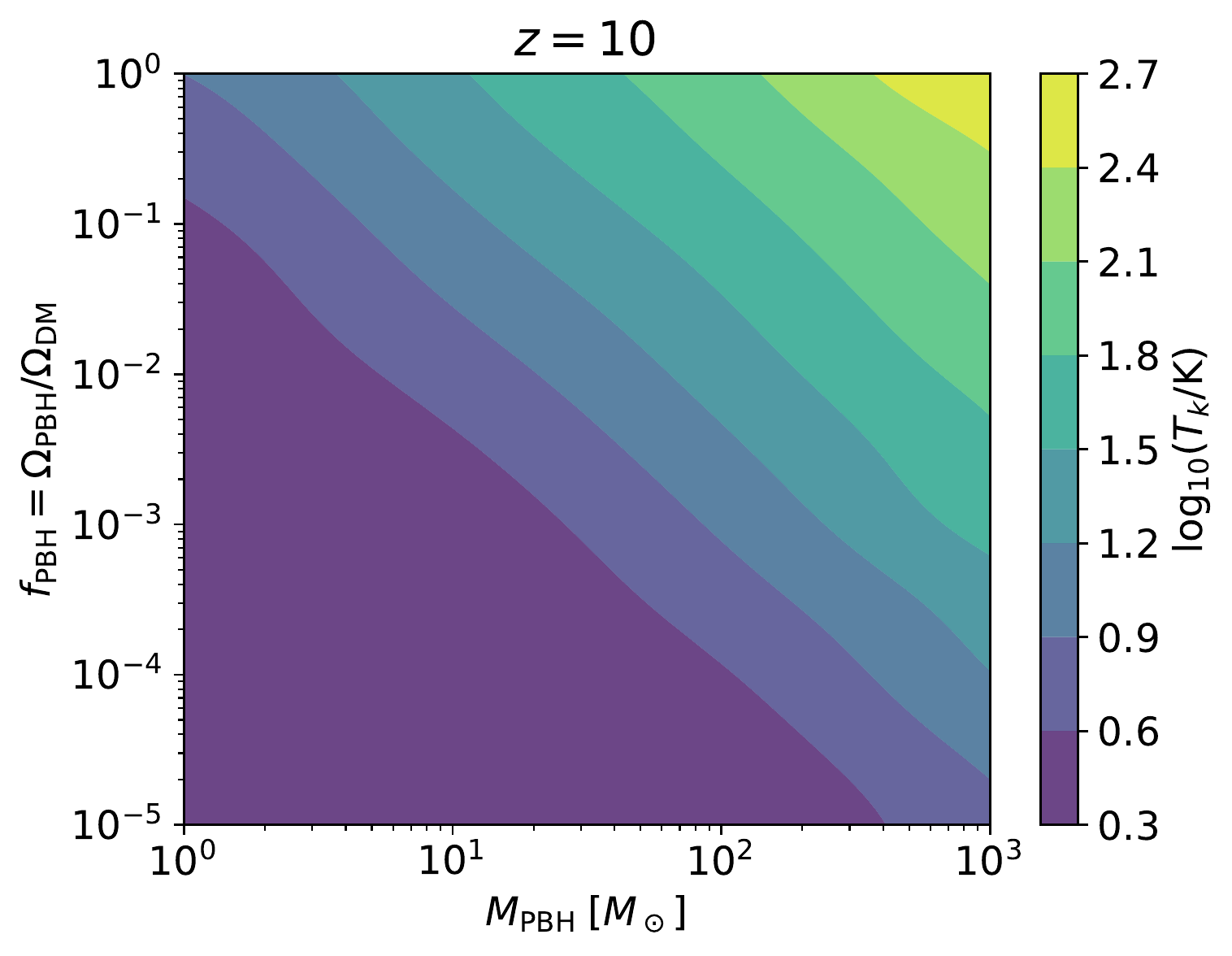}
 	\caption{Values of the IGM temperature at $z=10$ as a function of the mass $M_{\rm PBH}$ and fraction as DM $f_{\rm PBH}$ of PBHs.}
 	\label{fig:tempPBH}
 	\end{center}
 \end{figure}

The thermal histories of the IGM for the PBH scenarios are computed with the publicly available code {\tt 21cmFAST} \citep{Mesinger:2010ne}, properly modifying Eq.~(\ref{eq:heating}) to include the heating from PBH accretion\footnote{Shot noise effects are also included for completeness in the initial linear power spectrum, but their effects are negligible since they affect scales smaller than the resolution of the simulations, 1.5 Mpc.}. We start the simulations at $z=30$, when initial conditions of the gas temperature are computed with the recombination code {\tt CosmoRec} \citep{2011MNRAS.412..748C}, also modified to include accretion effects. We show the evolution of $T_k$ as a function of redshift for several PBH models in Fig.~\ref{fig:temperature}.  For now, we focus on the thin lines of the figure, which omit any astrophysical heating in our calculation except for the one from PBH accretion, to explicitly perceive the PBH effect. Standard astrophysical heating effects will be discussed below. Two cases for the PBH fraction are considered, $f_{\rm PBH} = 10^{-2}$ and $10^{-3}$, with three different PBH masses, $M_{\rm PBH} = 1, 10, $ and $ 10^2 M_\odot$. One can observe that larger PBH fractions give rise to a more pronounced heating effect and the IGM temperature remains higher compared to the case for the standard CDM model. With the PBH fraction fixed, PBHs with larger masses give rise to a higher temperature due to the larger mass accretion rate given by Eq.~(\ref{eq:BHrate}). Our calculation shows that the accretion effect becomes important if the PBH mass is $\gtrsim 10 M_\odot$ for the PBH fraction of $f_{\rm PBH} =10^{-2}$. However, the heating effect is not significant if the fraction is as small as $f_{\rm PBH} = 10^{-3}$ even for the PBH with mass as large as $10 M_\odot$.

To further understand the evolution of the IGM temperature and its dependence on the PBH parameters, we have run several {\tt 21cmFAST} simulations, spanning a grid where the PBH mass $M_{\rm PBH}$ varies from $1\, M_\odot$ and $10^3 M_\odot$, while $f_{\rm PBH}$ ranges from $10^{-5}$ to $1$. The resulting temperature is shown in Fig. (\ref{fig:tempPBH}). One can see that isotherms are determined by the condition
\begin{equation}
    f_{\rm PBH}M_{\rm PBH}^\beta = constant \, ,
\label{eq:fMacc}
\end{equation}
where $\beta \simeq 1.59$ is constant in redshift for $z \lesssim 30$. Hence, the gas temperature depends upon the joint quantity $f_{\rm PBH}M_{\rm PBH}^\beta$. Note that this is the expected scaling discussed in Sec. \ref{sec:acc}, taking into account the dependence of the luminosity with the PBH mass, with the power-law index $a=0.59$ from Tab. (\ref{table:adaf}). Below $f_{\rm PBH}M_{\rm PBH}^\beta \lesssim 10^{-1}$, the temperature saturates to the standard CDM value, which means that accretion effects are no longer relevant for such low masses and abundances. It is worth to emphasize that this is a different dependence than the present in the shot noise power spectrum, which scales as $f_{\rm PBH}M_{\rm PBH}$.

A more realistic treatment requires considering X-ray heating from different astrophysical sources, such as X-ray binaries. One could naively argue that the relevant quantity to constrain the abundance of PBHs is the difference between the number of absorbers in the PBH and the $\Lambda$CDM scenario, which would be both additionally suppressed in the same amount, and one thus would expect that X-ray heating would have little impact on this difference. However, as it will be explicitly shown, the number of absorbers can be significantly suppressed if astrophysical sources inject energetic radiation in the IGM, which affects our statistical approach based on the Poisson statistics, weakening thus the bounds. Actually, strong heating of the IGM can be a potential issue even in the standard $\Lambda$CDM scenario, worsening the prospects of observing the 21 cm forest \citep{2012MNRAS.425.2988M}.

There are many uncertainties regarding X-ray sources in the pre-Reionization universe. Compact X-ray binaries in starburst galaxies are the most likely origin of energetic radiation if their emissivity in the Cosmic Dawn is similar to the currently observed. Therefore, as usually done in the literature for the fiducial astrophysical heating, we extrapolate to our redshifts of interest the estimated luminosity from nearby X-ray binaries. The X-ray luminosity per star formation rate $L_X/{\rm SFR}$ integrated within the range $2-10$ keV from X-ray compact binaries lies between $\sim 10^{39}$ and $\sim 10^{40}$ erg s$^{-1}$  $M_\odot^{-1}$ yr between redshifts 0 and $\sim 4$ \citep{Madau:2016jbv}. We thus consider two limiting fiducial values of the luminosity per star formation rate, $10^{39}$ and $ 10^{40}$ erg s$^{-1}$  $M_\odot^{-1}$ yr, corresponding to mild and high X-ray heating scenarios, in order to illustrate the impact of astrophysical heating in our results. For the ratio of baryons into stars, $f_*$, we adopt a value of $f_*=0.01$, as suggested from radiation-hydrodynamic simulations of high-redshift galaxies (see, e.g., \cite{Wise:2014vwa}), or from the comparison of the star formation rate with the one derived from measurements of the UV luminosity function (see, e.g., \cite{Leite:2017ekt, Lidz:2018fqo} and references therein). The standard X-ray heating is already implemented in {\tt 21cmFAST}, see \cite{Mesinger:2010ne, Park:2018ljd} for more details. Figure (\ref{fig:temperature}) shows in thick lines examples of thermal histories for PBH scenarios including X-ray astrophysical heating with $L_X/{\rm SFR}= 10^{40}$ erg s$^{-1}$. Notice that, for that value, the onset of X-ray sources takes place at $z\sim 15$, being thus the astrophysical effects still very moderate at that epoch. They become however significant at $z \sim 10$. It is noteworthy that X-ray heating from astrophysical sources is more prominent for PBH scenarios with low accretion heating, increasing the temperature up to one order of magnitude in the limiting $\Lambda$CDM case. In contrast, the impact is reduced in PBH cases already substantially heated by accretion, increasing for instance by only a factor of $\simeq 2.5$ for $M_{\rm PBH}=10^2 \, M_\odot$ and $f_{\rm PBH}=10^{-2}$. Finally, the addition of the astrophysical X-ray heating does not change the dependence between $f_{\rm PBH}$ and $M_{\rm PBH}$ from Eq.~(\ref{eq:fMacc}) and shown in Fig.~(\ref{fig:tempPBH}).

\subsection{Number of absorber features}

\begin{figure}[t]
 \begin{center}
 	\includegraphics[width=0.48\textwidth]{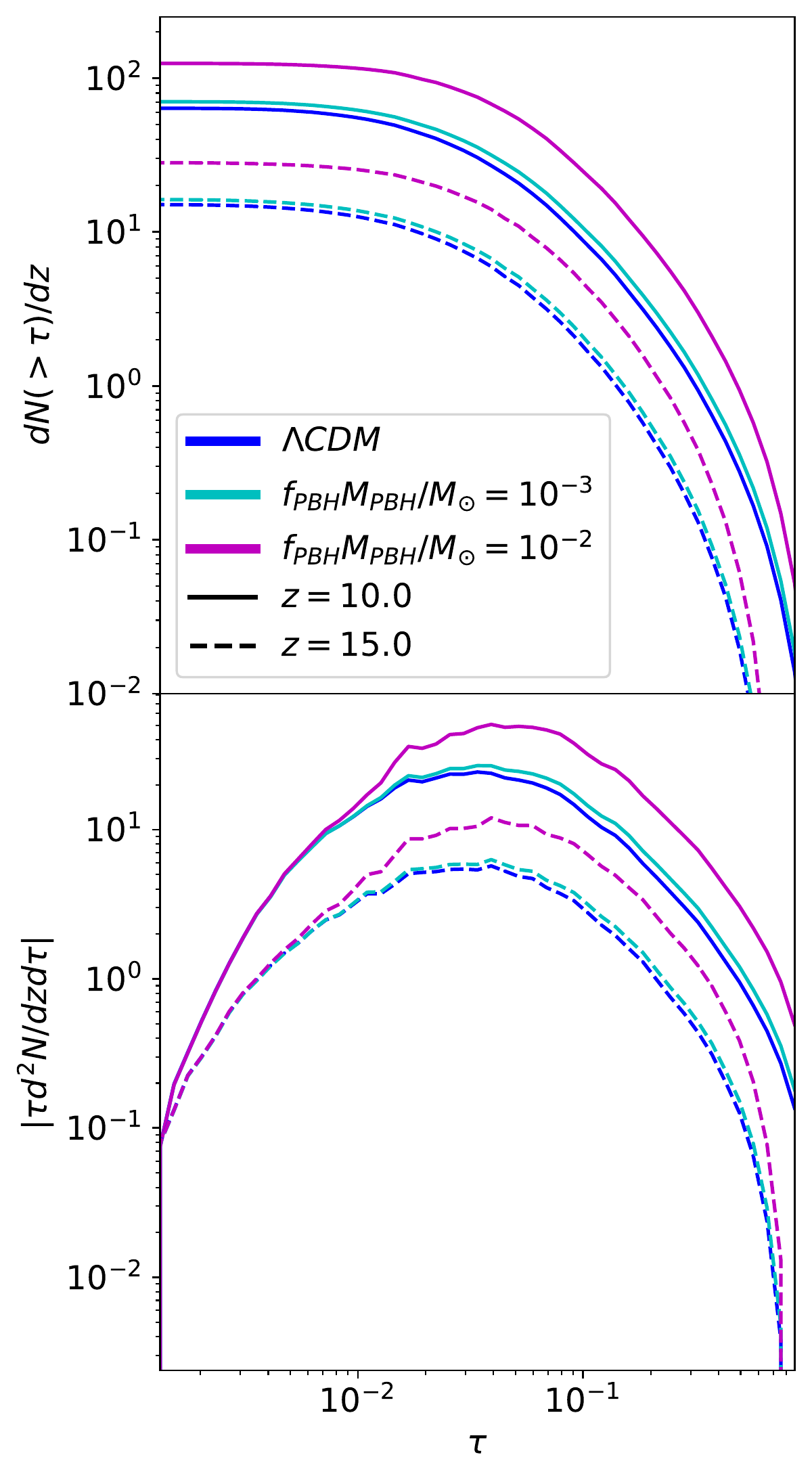}
 	\caption{Number of absorbers and its derivative neglecting the heating by accretion, for several values of $f_{\rm PBH} M_{\rm PBH}$ at two different redshifts. }
 	\label{fig:Absorbers_adiabatic}
 	\end{center}
 \end{figure}

\begin{figure*}[t]
 \begin{center}
 	\includegraphics[width=0.48\textwidth]{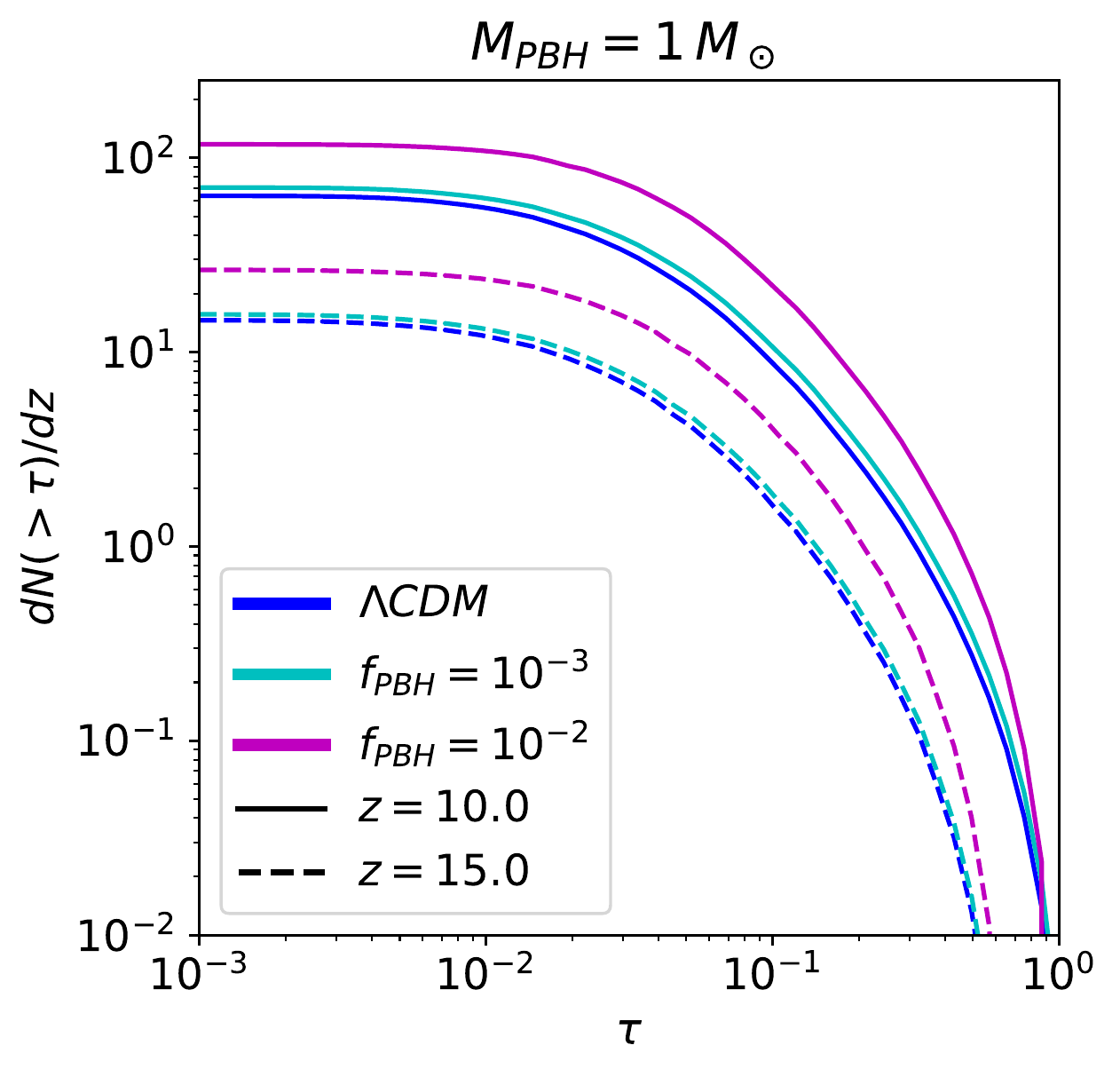}
 	\includegraphics[width=0.48\textwidth]{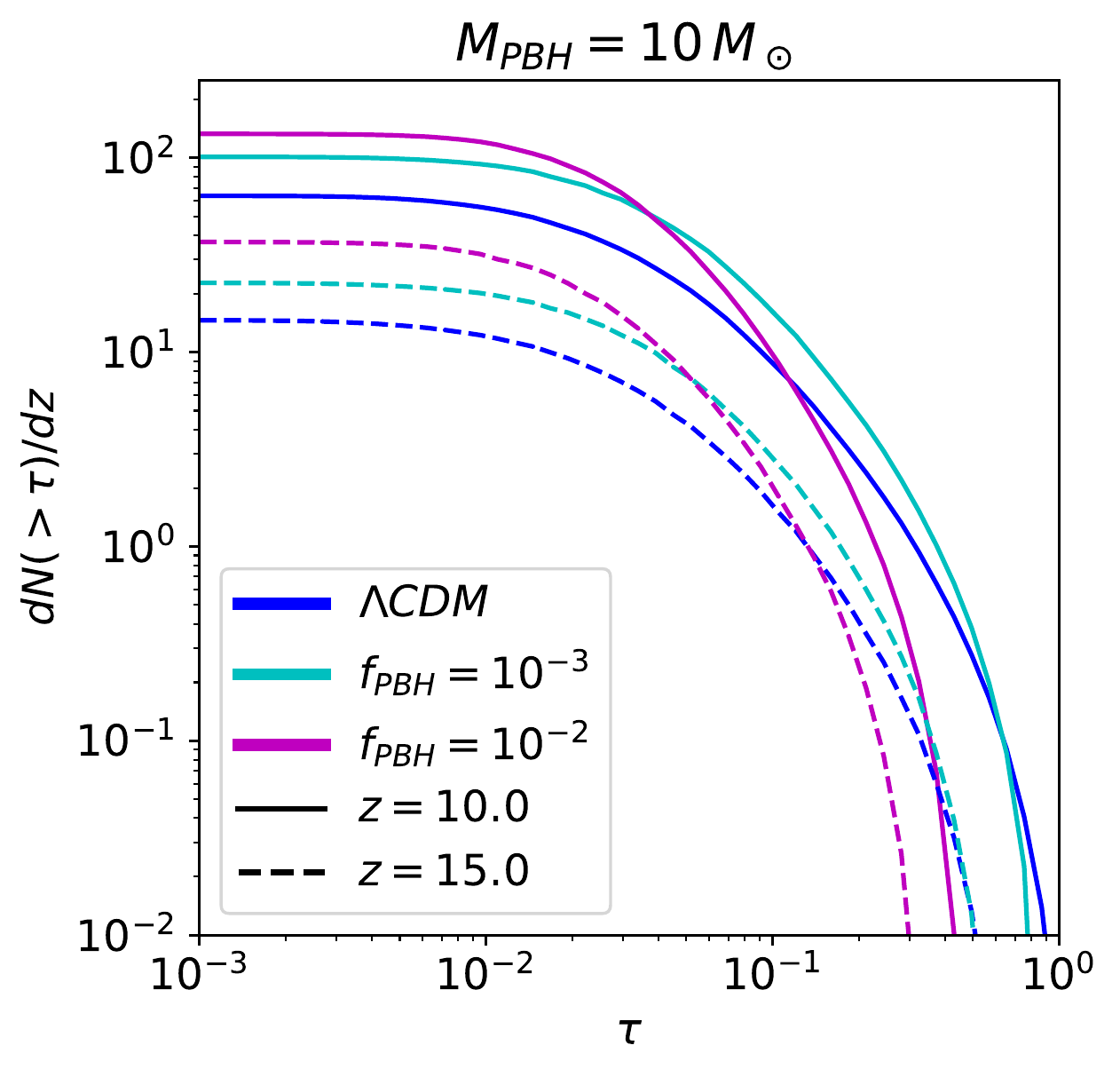}
 	\includegraphics[width=0.48\textwidth]{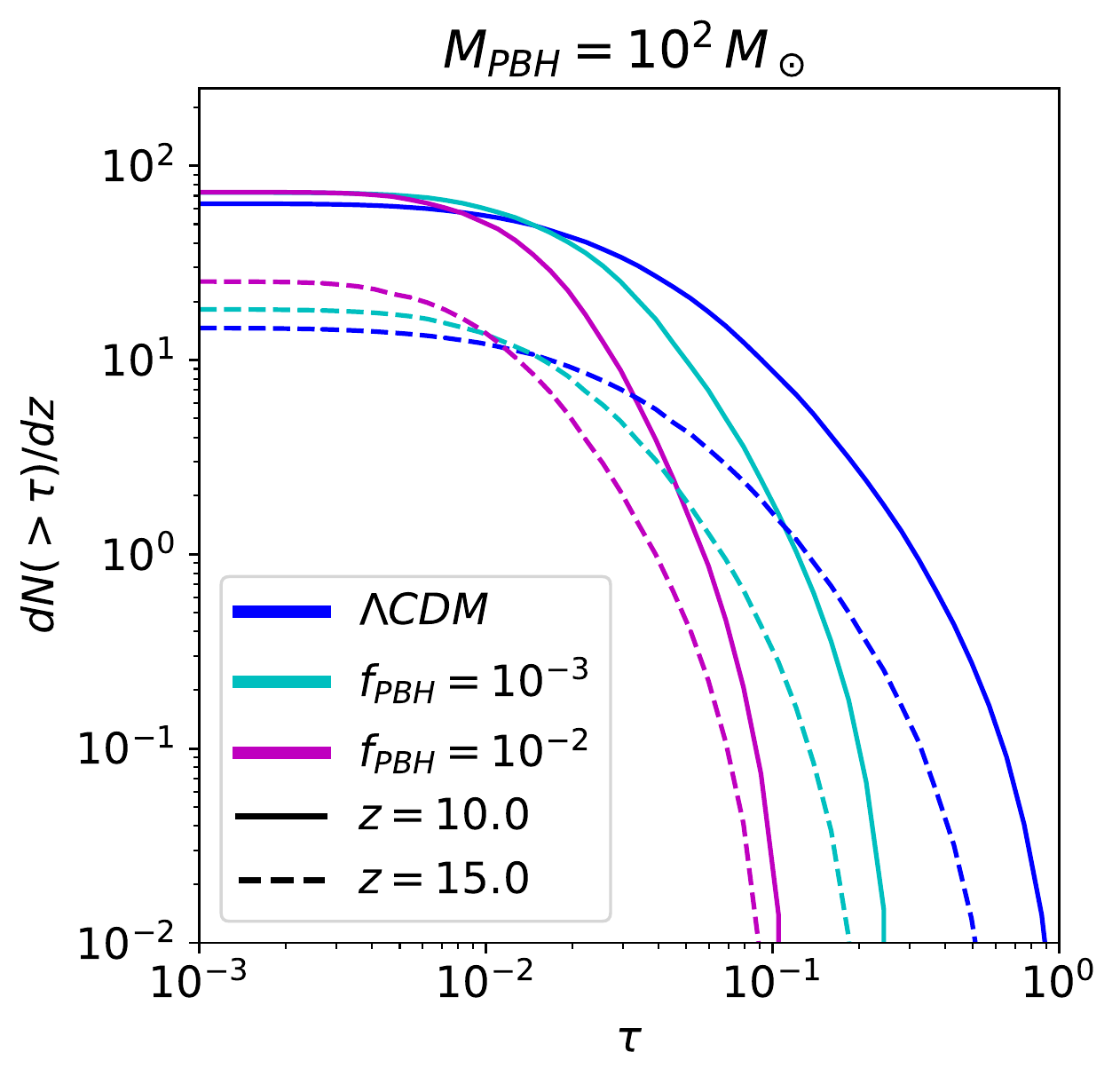}
 	\includegraphics[width=0.48\textwidth]{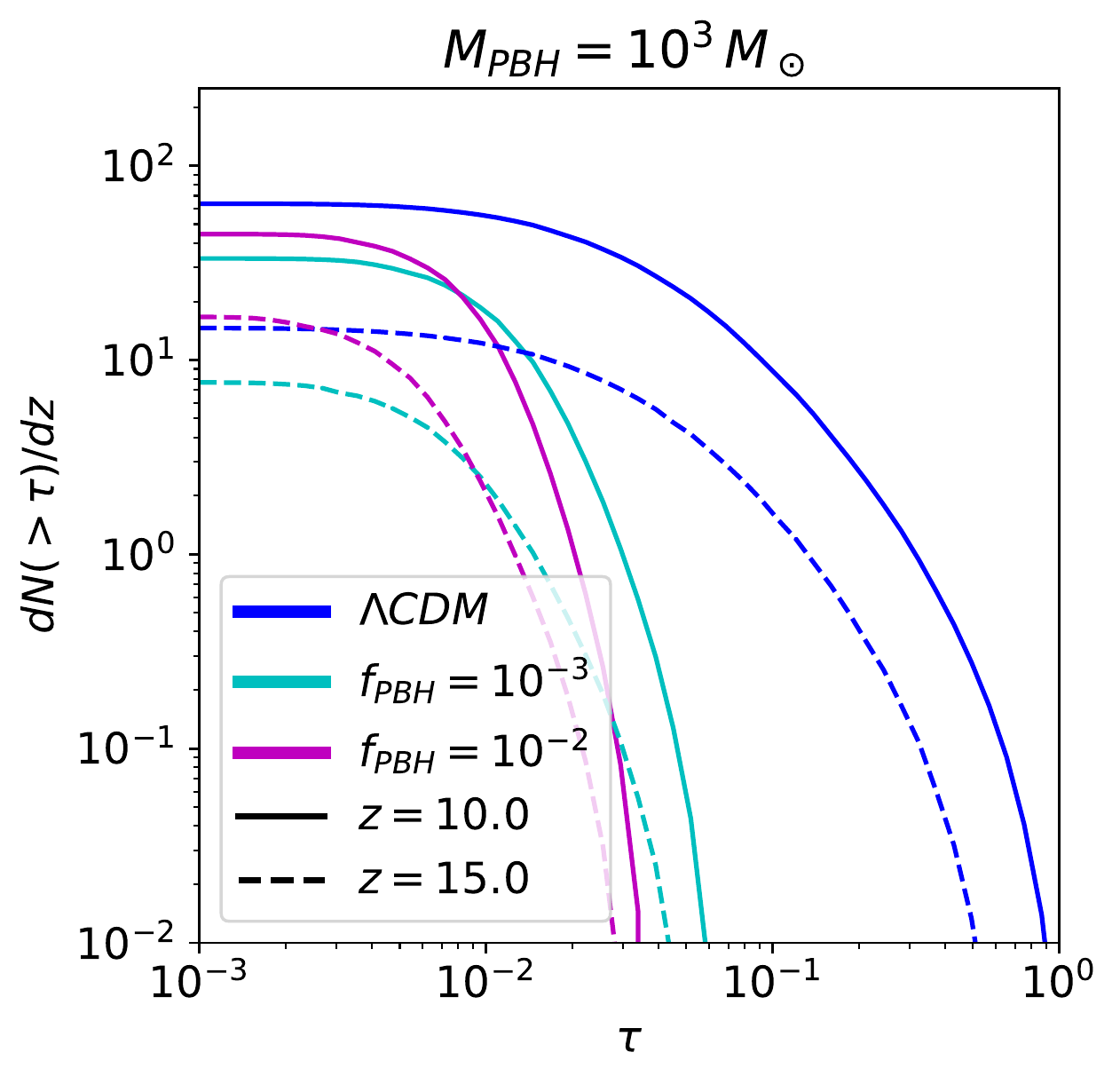}
 	\caption{Number of absorbers and its derivative for four values of $M_{\rm PBH}$ with an IGM heated only by accretion.}
 	\label{fig:Absorbers}
 	\end{center}
 \end{figure*}

With the thermal history of the IGM and the mass function of the DM halos, we calculate the expected number of absorption lines due to minihalos following the method presented in Sec.~\ref{sec:forest}. To investigate the effects of heating and the enhancement of the power spectrum by PBHs separately, we firstly focus on the shot noise contribution. Figure~(\ref{fig:Absorbers_adiabatic}) shows the number of absorption lines without the heating effect. Neglecting the heating by accretion (and by astrophysical heating), the change in the number of absorption lines due to minihalos simply reflects the change in the matter power spectrum. Because the power of the Poisson noise is proportional to $P\propto f_{\rm PBH}^2/n_{\rm PBH} \propto f_{\rm PBH}M_{\rm PBH}$, the effect depends on the combination of the parameters $f_{\rm PBH} M_{\rm PBH}$, and larger $f_{\rm PBH} M_{\rm PBH}$ results in the larger number of absorption features.  This behavior is shown in Fig.~(\ref{fig:Absorbers_adiabatic}) for the cumulative number of absorbers above a certain optical depth $\tau$, $dN(>\tau)/dz$ (as computed from Eq.~(\ref{eq:abundance1})), and its logarithmic derivative with respect to $\tau$, $\tau d^2N(>\tau)/dz/d\tau$, in the top and bottom panel respectively. Both quantities reflect an enhancement powered by $f_{\rm PBH}M_{\rm PBH}$, which is approximately equally distributed along $\tau$.

We now include the heating effect from PBH accretion, using the IGM temperature instead of the adiabatic one for computing the Jeans mass. Examples for four values of the PBH masses are shown in Fig.~(\ref{fig:Absorbers}). 
Because the IGM temperature scales with the parameters $f_{\rm PBH}$ and $M_{\rm PBH}$ differently than the shot noise contribution, the degeneracy between these two parameters found in the power spectrum is broken. The increase in the Jeans mass reduces the range of mass integration in Eq.~(\ref{eq:mass_integration}), and therefore it counteracts the enhancement effect due to the additional contribution of the power spectrum. As can be seen in the right panel of Fig. (\ref{fig:taurho}), the highest optical depths appear in low-mass minihalos, while the largest minihalos produce lower optical depths. Increasing the IGM temperature and the Jeans mass reduces the low-mass halos, where the highest values of $\tau$ are obtained. Hence, the accretion effect is more pronounced at large optical depths.

Comparing the first panel of Fig.~(\ref{fig:Absorbers}) with Fig.~(\ref{fig:Absorbers_adiabatic}), one can see that the accretion effect is mostly negligible at $M_{\rm PBH}= M_\odot$ for the abundances assumed. At PBH masses $\lesssim 10 M_\odot$, for relatively high fractions $f_{\rm PBH} \simeq 10^{-2}$, the enhancement is still noticeable, especially for higher redshifts. Nonetheless, at $M_{\rm PBH} = 10 M_\odot$, the largest optical depths are already suppressed due to the accretion effect. For the cases with $M_{\rm PBH}=10^2 M_\odot$, the number of absorption lines only slightly increases compared to the standard $\Lambda$CDM  case (being more prominent at $z=15$). On the other hand, the increase in the Jeans mass due to the extra heating overcomes the enhancement effect in the matter power spectrum, suppressing the number of absorption lines with respect to the CDM case, having the opposite effect. For $M_{\rm PBH}=10^3 M_\odot$, the accretion effect dominates, suppressing the number of absorbers along the $\tau$ range (except for $f_{\rm PBH}=10^{-2}$ at $z=15$, which still shows a small enhancement at very low optical depths).

One can also compute the cumulative number of absorption features above a certain optical depth $\tau$ at redshift $z$ within a redshift interval $\Delta z$, which can be obtained by integrating Eq.~(\ref{eq:abundance1}) over redshift. While in principle one should integrate down to $z=0$ to account for all the absorption features in the forest along the line of sight, it is expected that neutral minihalos would be extinct after Reionization at $z\sim 6$. Given the uncertainties in the number of minihalos when Reionization is in progress, as a conservative assumption, we limit our redshift integration to an interval $\Delta z=1$, where $dN(>\tau)/dz$ can be approximated as constant. The cumulative number of absorbers can thus be written as
\begin{equation}
N(>\tau,z) = \int_{z-\Delta z}^{z} dz' \frac{dN(>\tau)}{dz'} \simeq \frac{dN(>\tau)}{dz} \Delta z.
\label{eq:nums}
\end{equation}
In the formula above, the minimum optical depth is determined by the sensitivity of the experiment. Given Fig. (\ref{fig:Absorbers}) and the sensitivity discussion from Eq.~(\ref{eq:Smin}), one can infer that optical depths of $\mathcal{O}(10^{-2})$ would be sufficient for bright enough radio sources. For concreteness, we split our computations between the ranges within $\tau=0.01$ and $\tau=0.03$, and above $\tau=0.03$, to test how robust are the results to different choices of the threshold optical depth, and how relevant are the distinct ranges of $\tau$. Table (\ref{table:numbers}) shows some examples of the number of absorbers at $z=10$ for different PBH scenarios, for $N(\tau>0.03)$ and for $N(0.01\leq\tau\leq0.03) = N(\tau>0.01) - N(\tau>0.03)$.

\subsection{Bounds on the PBH abundance}

\begin{table}
	\begin{center}
		\begin{tabular}{|c|c|c|}
			\hline
			$(M_{\rm PBH},f_{\rm PBH})$ & $N(0.01\leq\tau\leq0.03)$ & $N(\tau>0.03)$ \\ 
			\hline\hline
			$(1,10^{-2})$ & 34 & 74 \\
			\hline
			$(1,10^{-3})$& 23 & 39 \\
			\hline
			$(10,10^{-2})$ & 54 & 65 \\
			\hline
			$(10,10^{-3})$& 32 & 61 \\
			\hline
			$(100,10^{-2})$ & 42 & 9 \\
			\hline
			$(100,10^{-3})$& 35 & 25 \\
			\hline
			$\Lambda$CDM $(0,0)$& 22 & 33 \\
			\hline
		\end{tabular}
	\end{center}
	\caption{Number of absorbers at $z=10$ computed using Eq. (\ref{eq:nums}) assuming only heating by PBH accretion.}
	\label{table:numbers}
\end{table}

\begin{figure*}[t]
 \begin{center}
    \includegraphics[width=0.49\textwidth]{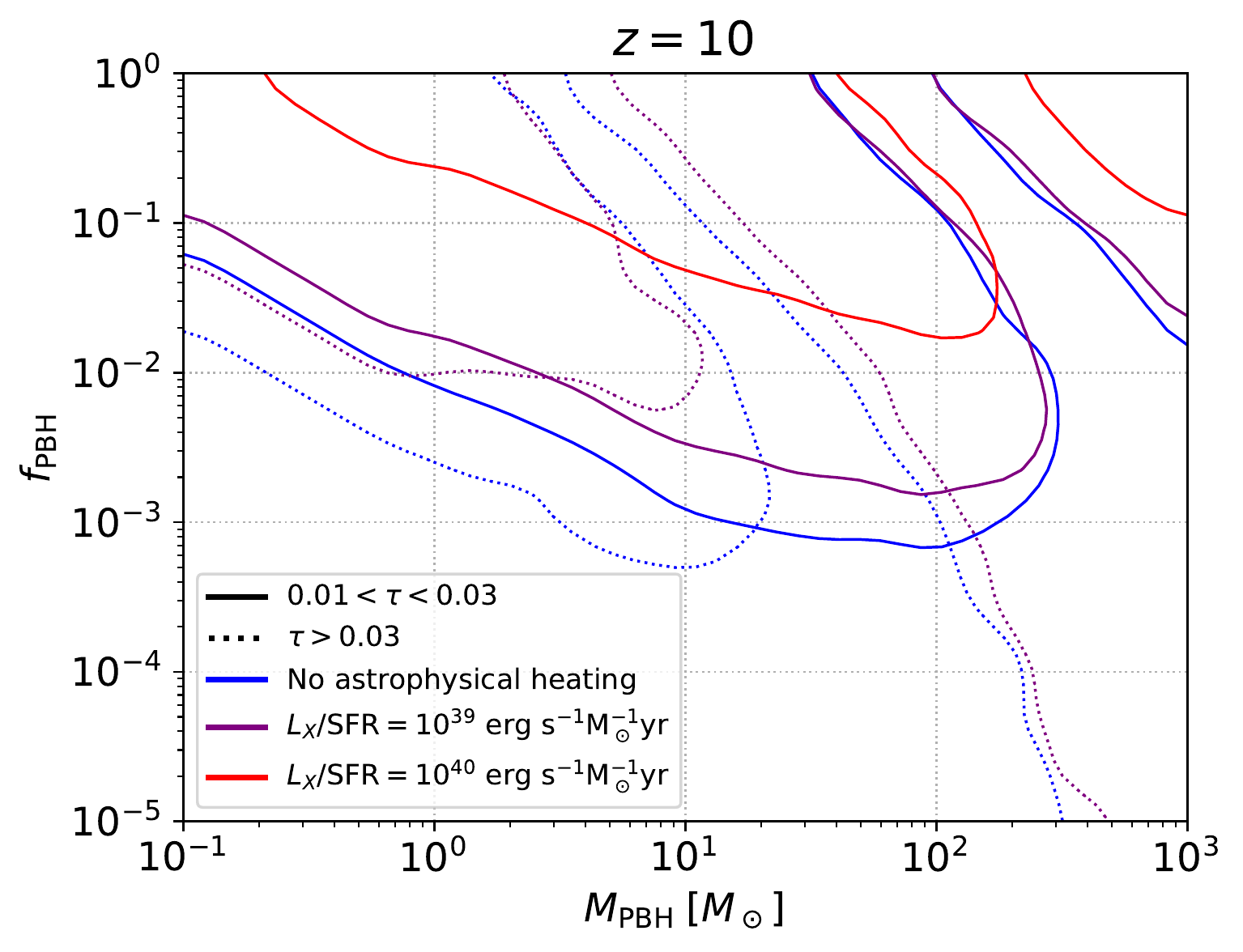}
 	\includegraphics[width=0.49\textwidth]{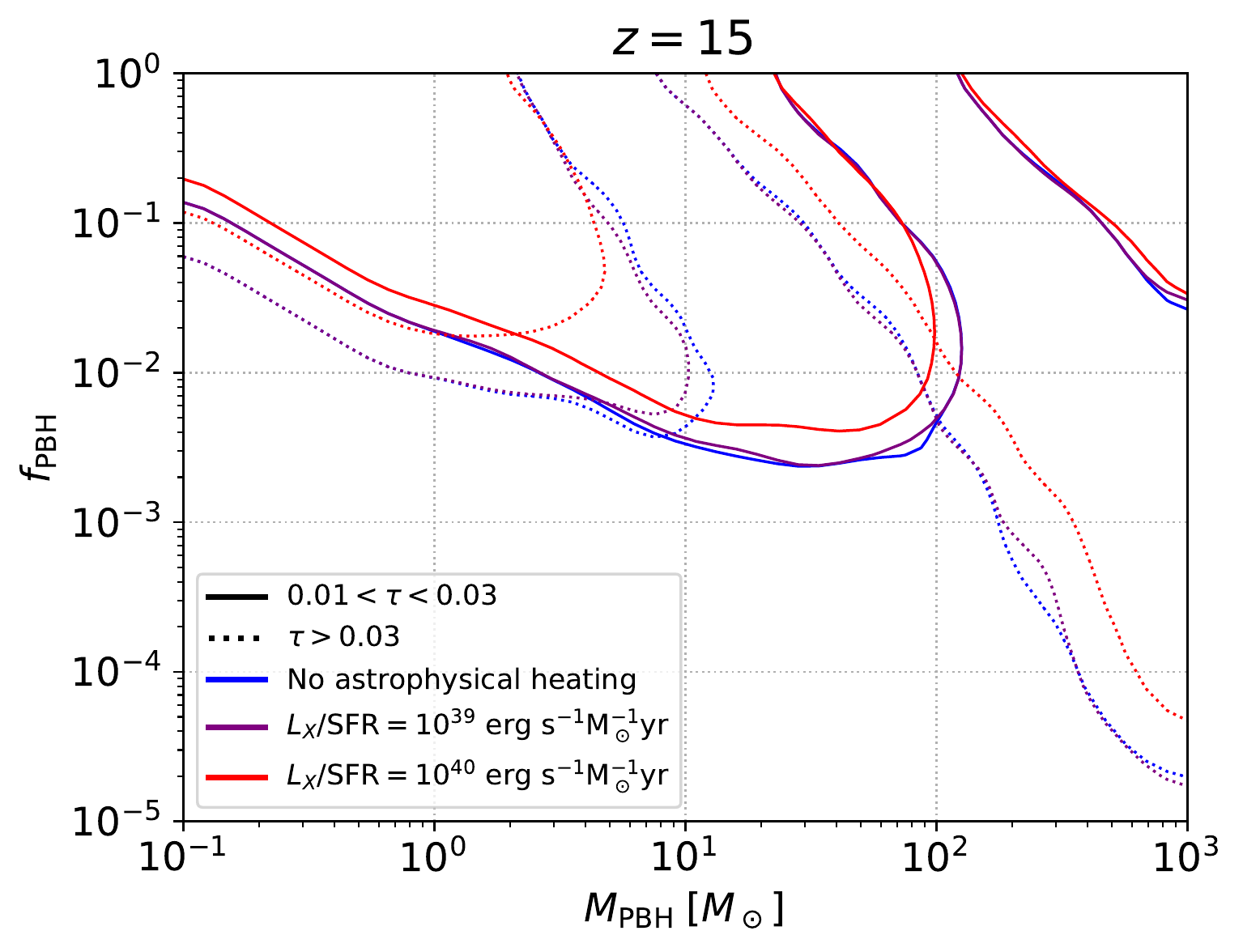}
 	\caption{Constraints on the abundance of PBHs as a function of their mass applying the criterion form Eq. (\ref{eq:condition}), for three astrophysical heating regimes and a background source located at $z=10$ (left) and $z=15$ (right). Solid line accounts for $N(0.01\leq\tau\leq0.03)$, while dotted line for $N(\tau>0.03)$.}
 	\label{fig:bounds_heating}
 	\end{center}
 \end{figure*}

 \begin{figure*}[t]
 \begin{center}
 	\includegraphics[width=0.99\textwidth]{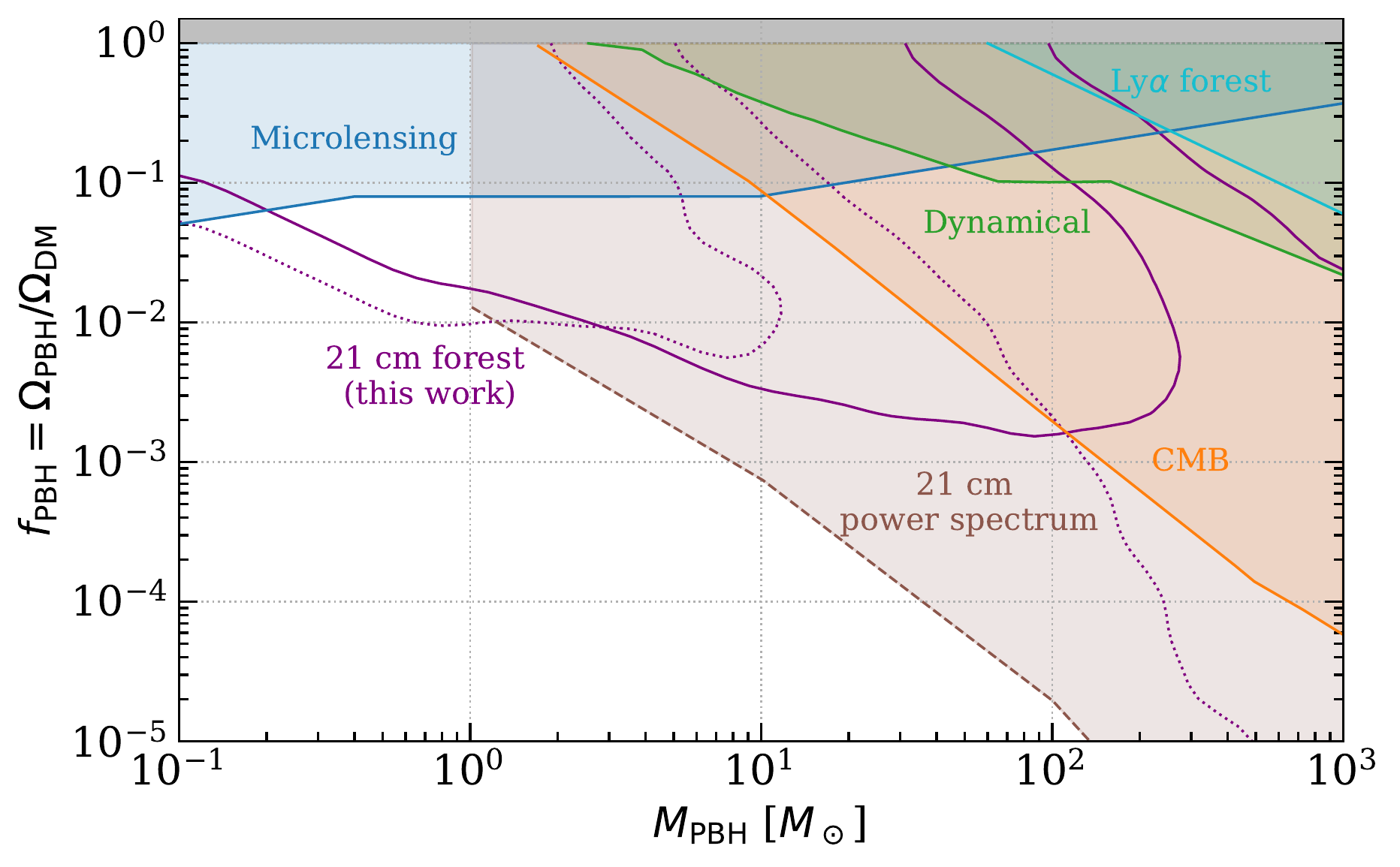}
 	\caption{Constraints on the abundance of PBHs as a function of their mass. Purple lines denote the prospects from our analysis, employing the number of absorbers of the 21 cm forest at $z=10$ with mild astrophysical heating, applying the criterion from Eq. (\ref{eq:condition}). Solid line accounts for $N(0.01\leq\tau\leq0.03)$, while dotted line for $N(\tau>0.03)$. For comparison, current bounds from other probes have been included, namely: non-observation of microlensing events (blue) from the MACHO \citep{2001ApJ...550L.169A}, EROS \citep{2007A&A...469..387T}, Kepler \citep{Griest:2013aaa} and Icarus \citep{Oguri:2017ock}  collaborations; PBH accretion signatures on the CMB (orange), assuming disk ADAF accretion \citep{Poulin:2017bwe}; dynamical constraints, such as disruption of stellar systems by the presence of PBHs (green), on wide binaries \citep{2014ApJ...790..159M} and on ultra-faint dwarf galaxies \citep{2016ApJ...824L..31B}; power spectrum from the Ly$\alpha$ forest (cyan) \citep{2019PhRvL.123g1102M}. Dotted brown line corresponds to forecasts for SKA from the impact of accretion on the 21 cm power spectrum of the IGM \citep{Mena:2019nhm}. Figure created with the publicly available \texttt{Python} code \texttt{PBHbounds}\protect\footnotemark     \citep{bradley_j_kavanagh_2019_3538999}. }
 	\label{fig:bounds}
 	\end{center}
 \end{figure*}

The aforementioned results allow us to explore the possible bounds on the population of PBHs that could be achieved by observing the absorption features from the 21 cm forest. We denote the number of absorbers within a given optical depth range as $N_{i}$, with $i={\rm \Lambda CDM}, {\rm PBH}$ for the standard and PBH case respectively, with an uncertainty given by $\Delta N_i$. Assuming Poisson statistics, where the variance equals the mean, one can write $\Delta N_i = \sqrt{N_i}$ for the 1$\sigma$ uncertainty interval. We thus consider that 21 cm forest observations may be able to distinguish between the $\Lambda$CDM case and a PBH scenario if $N_{\rm \Lambda CDM}+\Delta N_{\rm \Lambda CDM} < N_{\rm PBH}-\Delta N_{\rm PBH}$ if the number of absorbers is enhanced, or $N_{\rm \Lambda CDM}-\Delta N_{\rm \Lambda CDM} > N_{\rm PBH}+\Delta N_{\rm PBH}$ if the accretion effects suppress the absorption features, and their number is reduced. Both conditions can be summarized as\footnote{Alternatively, one could also employ $\sqrt{\Delta N^2_{\Lambda CDM} + \Delta N^2_{\rm PBH}}$ instead of $\Delta N_{\rm \Lambda CDM} + \Delta N_{\rm PBH} $, but since $\Delta N_i$ are positive by definition, and the latter is always equal or larger than the former, we opt for Eq.~(\ref{eq:condition}) for being more conservative.}
\begin{equation}
    | N_{\rm PBH}-N_{\rm \Lambda CDM} | > \Delta N_{\rm \Lambda CDM} + \Delta N_{\rm PBH} 
    \label{eq:condition}
\end{equation}

Figure (\ref{fig:bounds_heating}) shows a forecast of the upper bounds on the fraction of PBHs as DM as a function of their mass, based on the criterion from Eq.~(\ref{eq:condition}) for a hypothetical background source located at $z=10$ (left) and $z=15$ (right). The solid lines correspond to $N(0.01\leq\tau\leq0.03)$, while the dotted ones to $N(\tau>0.03)$. Three heating scenarios are considered, one neglecting astrophysical X-ray heating, considering only heating by accretion (blue lines), and two fiducial values of the X-ray luminosity from X-ray binaries, accounting for mild ($L_X/{\rm SFR}= 10^{39}$ erg s$^{-1}$, purple lines) and high ($L_X/{\rm SFR}= 10^{40}$ erg s$^{-1}$, red lines) heating, as discussed in Sec. \ref{sec:temperatureIGM}. 

It is straightforward to understand the shape of these forecasted limits. Focusing now in the range $0.01\leq\tau\leq0.03$ (solid lines), one can distinguish two different regimes on the expected bounds, roughly the same at both redsfhits. At low masses, $M_{\rm PBH} \lesssim 100$ $M_\odot$, and relatively low PBH fractions, $f_{\rm PBH}<10^{-1}$, the shot noise effect dominates, and allows to probe that region of the parameter space. Given that the enhancement is powered by the product $f_{\rm PBH}M_{\rm PBH}$, the upper limit follows a line with the dependence $f_{\rm PBH} \propto M_{\rm PBH}^{-1}$. In the other extreme, at high masses, $M_{\rm PBH} \gtrsim 100$ $M_\odot$, and large PBH fractions, $f_{\rm PBH}>10^{-2}$, the number of absorbers is suppressed due to heating from accretion. This constraint depends upon the IGM temperature, which recalling from Sec. \ref{sec:temperatureIGM}, depends upon the joint product $f_{\rm PBH} M_{\rm PBH}^{\beta}$, with $\beta \simeq 1.59$. 
Hence, the shape of the upper limit reflects the isotherms of Eq.~(\ref{eq:fMacc}), shown in Fig. (\ref{fig:tempPBH}), scaling as $f_{\rm PBH} \propto M_{\rm PBH}^{-\beta}$. 

Amidst these two regimes, both effects compensate, leading to a number of absorbers similar to those expected from the $\Lambda$CDM case. This trade off leaves a window around $\sim 100 M_\odot$ where the possible bounds relax and even disappear. The analysis employing $N(\tau>0.03)$ is qualitatively the same, albeit the ranges of masses delimiting the different regimes change. Larger optical depths are more sensitive to the drop in the number of minihalos due to the IGM heating, as is manifest in Fig. (\ref{fig:Absorbers}), and thus the accretion regime starts at a lower mass, roughly $\sim 10 M_\odot$ in this case. Note that measuring at different ranges of $\tau$ can probe distinct regions of the parameter space, and combining them, it would be possible to remove the windows with no bounds.

While the aforementioned dependences in Fig. (\ref{fig:bounds_heating}) behave similarly at $z=10$ and $z=15$, astrophysical heating establishes a key difference between both redshifts. At $z=15$, X-ray binaries and other sources are not abundant enough to greatly impact the IGM temperature, as seen in Fig. (\ref{fig:temperature}), and the bounds are not significantly modified for large astrophysical heating. However, at $z=10$, X-ray background radiation could be important enough to substantially increase the IGM temperature, leaving a deep impact in our analysis. While the limits only worsen by a factor of a few in the mild heating scenario, the case with large X-ray luminosity reduces the bounds by more than an order of magnitude. Moreover, this weakening is not equal for all the ranges of optical depths, disappearing completely for $\tau>0.03$. These facts show the great dependence of the 21 cm forest results upon the IGM heating, becoming mandatory to properly understand the growth of X-ray sources during the Cosmic Dawn in order to infer robust bounds on PBHs.

In order to better understand the strength of these forecasted bounds, it is useful to compare them with the constraints from other studies. Figure (\ref{fig:bounds}) depicts a sample of our bounds together with the current constraints on the abundance on this range of masses extracted from different probes. The purple lines account for the case $z=10$ and mild heating ($L_X/{\rm SFR}= 10^{39}$ erg s$^{-1}$), for both ranges of $\tau$ considered in Fig. (\ref{fig:bounds_heating}). We opt to show this scenario since it is the intermediate case at $z=10$, and roughly coincides with the bounds at $z=15$ for mild and vanishing X-ray sources, being thus a representative example. The considered experiments include the non-observation of microlensing events (blue) from the MACHO \citep{2001ApJ...550L.169A}, EROS \citep{2007A&A...469..387T}, Kepler \citep{Griest:2013aaa} and Icarus \citep{Oguri:2017ock}  collaborations; PBH accretion signatures on the CMB anisotropies with Planck data (orange) \citep{Poulin:2017bwe}; dynamical constraints, such as disruption of stellar systems by the presence of PBHs (green), on wide binaries \citep{2014ApJ...790..159M} and on ultra-faint dwarf galaxies \citep{2016ApJ...824L..31B}; and the power spectrum from the Ly$\alpha$ forest enhanced by shot noise (cyan) \citep{2019PhRvL.123g1102M}. Forecasted limits from the imprints of accretion in the IGM 21 cm power spectrum from future observations with SKA are also shown (dotted brown line) \citep{Mena:2019nhm}.

\footnotetext{\url{https://github.com/bradkav/PBHbounds}}

One can see that our estimate from future 21 cm forest observations at $z\sim 10$ may be able to explore a broad region in the parameter space not covered yet by current probes, being sensitive to even small abundances of PBHs, and partly overlapping the prospects from the 21 cm power spectrum of the IGM. Note that Ly$\alpha$ forest bounds present the same scaling with $M_{\rm PBH}$ than our predictions at masses $\lesssim 1 M_\odot$. It is because both are based on the contribution from the shot noise power spectrum. However, the 21 cm forest may provide much more stringent constraints since it can probe smaller scales at higher redshifts. Similarly, both the 21 cm power spectrum prospects and the CMB bounds scale similar to those from the accretion effect obtained here, because both bounds would depend on the IGM temperature, heated in the same way due to ADAF disk accretion. Other CMB bounds have been derived in the literature modifying the accretion modeling, e.g., assuming spherical accretion of PBHs within halos \citep{2020PhRvR...2b3204S}, which present a different scaling. Finally, it is worth noting that observations of BH merger events from gravitational waves have been claimed to set stringent bounds in the range of masses from $10^{-2}$ to $\sim 10^{2} M_\odot$, by demanding that the predicted merger rates of PBH binaries cannot exceed the measured ones. This can be achieved either employing individual merger events \citep{2018PhRvD..98b3536K, 2019PhRvD.100b4017A} or searches of a stochastic background of gravitational waves  \citep{2020JCAP...08..039C}. However, these limits assume PBHs as Schwarzschild BHs, while it should be more appropriate to use cosmological BH solutions embedded in the FLRW metric. Properly accounting for a cosmological metric may allow to avoid those bounds, getting merger rates compatible with the LIGO and VIRGO detections \citep{Boehm:2020jwd}. For that reason, we exclude these bounds in Fig.~(\ref{fig:bounds}).

\section{Conclusions}
\label{sec:conclusions}

In this article, we have investigated the effects of PBHs on the 21 cm forest due to minihalos in the Epoch of Reionization and Cosmic Dawn. The 21 cm forest presents several advantages over other approaches to probe the IGM. It allows studying earlier periods than the Ly$\alpha$ forest, previous to Reionization. Moreover, unlike the 21 cm signal from the IGM, it avoids the problem of foreground removal since backlight radio sources would be much brighter. Additionally, arbitrary small scales can be probed, relying on the frequency resolution. However, it also suffers from two main uncertainties and drawbacks. 

On the one hand, it depends upon the existence of bright enough radio sources at high redshift, none of which has been observed yet, although there are good prospects for discovering them in the future. On the other hand, it is strongly dependent on the heating state of the IGM. Since the number of minihalos is limited by the Jeans mass, a too hot gas would drastically reduce the number of absorbers, eventually spoiling the statistical power.

The presence of PBHs as DM may alter the number of 21 cm absorption lines in two ways. One is increasing their number through enhancing the mass function due to the newly induced Poisson shot-noise isocurvature mode on small scales. The other is further heating the IGM due to the gas accretion onto the PBHs, making the Jeans mass scale larger, reducing the number of absorbers. We estimate via a semianalytic approach the number of absorption lines for PBH masses of $0.1\leq M_{\rm PBH}/M_\odot \leq 10^4$ and abundances $10^{-5} \leq f_{\rm PBH} \leq 1$. We find that the PBHs have a negligible impact on the number of absorption lines for $f_{\rm PBH} \lesssim 10^{-3}$ for their masses up to $100 M_\odot$. If the mass of PBHs is smaller than $\lesssim 100 M_\odot$, the heating effect is weak, and the number of absorption lines monotonically increases as $f_{\rm PBH}$ increases. If the mass of PBHs is as large as $100 M_\odot$, the heating effect can be significant, and the number of absorption lines becomes smaller compared to the standard $\Lambda$CDM case. However, this behavior depends on the range of optical depths considered. Absorbers with large values of $\tau$ are more affected by the suppression due to accretion than those with lower values. Hence the bounds on intervals of large optical depths are dominated by the accretion heating effect at large PBH masses.

We find that our results strongly rely on the assumed astrophysical X-ray heating. A too hot IGM can substantially reduce the number of absorbers in both PBH and CDM scenarios, weakening considerably the obtained bounds and even preventing us from reaching the statistical sensitivity to discriminate between both models if large values of the optical depth are employed. However, mild heating scenarios may only feebly reduce the limits. Nevertheless, the presence and luminosity of X-ray sources during the Cosmic Dawn is quite unclear. It is customary to assume similar luminosities at high redshift to those existing in the nearby universe. However, the EDGES collaboration claimed a detection of the 21 cm global signal at $z \simeq 17$, which, if it is verified, would imply a cold IGM at $z >10$ \citep{2018Natur.555...67B}. It is therefore not clear how hot the IGM may be, and further studies are mandatory for a proper determination of the possible bounds on the PBH abundance.

Our approach is based on a semianalytic framework, mainly following the formalism from \cite{2002ApJ...579....1F}, which significantly simplifies the extremely complex physics of the IGM. The virial characterization of minihalos, the computation of the gas density profile or neglecting the Ly$\alpha$ coupling to compute the spin temperature inside minihalos are well motivated although oversimplistic approximations, which may have an impact on the final bounds. In order to obtain more reliable results, hydrodynamical simulations would be required. However, we have taken several conservative assumptions along our work to ensure the robustness of the results. For instance, for a background source located at redshift $z$, we have employed a range of integration of $\Delta z = 1$ in Eq.~(\ref{eq:nums}), while in general, there should be contributions from lower redshifts $<z$, as long as neutral minihalos still exist. In addition, the Jeans mass has been taken as the minimal threshold for considering minihalos. This could be seen as a conservative hypothesis, since minihalos with lower masses formed previously in a colder IGM are thus excluded, which could contribute to the number of absortion features \citep{2011MNRAS.417.1480M}. Relaxing some of these conditions may imply stronger bounds than the obtained here.

The statistical test based on the criterion from Eq.~(\ref{eq:condition}) based on the Poisson statistics is considerably simple. Nevertheless, it suffices for providing a rough conservative estimate of the expected bounds, rather than a precise evaluation of the limits. Further rigorous approaches, involving Fisher matrix analyses or Monte Carlo methods, should be employed to ensure more robust constraints. Nonetheless, the approach followed here is enough for illustrating the power of the 21 cm forest for constraining the abundance of PBHs as DM.

%%% contents

%\begin{ack}
\section{Acknowledgments}
PVD thanks the Nagoya University for its hospitality during the early stages of this work. We thank Masahiro Takada for insightful conversations in the first stages of the work, and Kyungjin Ahn,  Kenji Hasegawa, Hayato Shimabukuro and Kenji Kadota for helpful discussions. We also thank Samuel J. Witte for his modified version of {\tt CosmoRec}. This work was in part supported by the JSPS grant numbers 18K03616, 17H01110 and JST AIP Acceleration Research Grant JP20317829 and FOREST Program JPMJFR20352935. The work is also supported  by JSPS Core- to-Core Program (JPJSCCA20200002). PVD is supported by the Spanish MINECO grants SEV-2014-0398 and FPA2017-85985-P, by the Generalitat Valenciana grant PROMETEO/2019/083, and by the European Union’s Horizon 2020 research and innovation program under the Marie Skłodowska-Curie grant agreements No. 690575 (InvisiblesPlus) and 674896 (Elusives). We have made use of the Python package {\tt Colossus} \citep{2018ApJS..239...35D}.
%\end{ack}

\bibliographystyle{aasjournal}
\bibliography{biblio}

\begin{thebibliography}{}
\expandafter\ifx\csname natexlab\endcsname\relax\def\natexlab#1{#1}\fi
\providecommand{\url}[1]{\href{#1}{#1}}
\providecommand{\dodoi}[1]{doi:~\href{http://doi.org/#1}{\nolinkurl{#1}}}
\providecommand{\doeprint}[1]{\href{http://ascl.net/#1}{\nolinkurl{http://ascl.net/#1}}}
\providecommand{\doarXiv}[1]{\href{https://arxiv.org/abs/#1}{\nolinkurl{https://arxiv.org/abs/#1}}}

\bibitem[{{Abbott} {et~al.}(2016){Abbott}, {Abbott}, {Abbott}, {Abernathy},
  {Acernese}, {Ackley}, {Adams}, {Adams}, {Addesso}, {Adhikari}, {Adya},
  {Affeldt}, {Agathos}, {Agatsuma}, {Aggarwal}, {Aguiar}, {Aiello}, {Ain},
  {Ajith}, {Allen}, {Allocca}, {Altin}, {Anderson}, {Anderson}, {Arai},
  {Araya}, {Arceneaux}, {Areeda}, {Arnaud}, {Arun}, {Ascenzi}, {Ashton}, {Ast},
  {Aston}, {Astone}, {Aufmuth}, {Aulbert}, {Babak}, {Bacon}, {Bader}, {Baker},
  {Baldaccini}, {Ballardin}, {Ballmer}, {Barayoga}, {Barclay}, {Barish},
  {Barker}, {Barone}, {Barr}, {Barsotti}, {Barsuglia}, {Barta}, {Bartlett},
  {Bartos}, {Bassiri}, {Basti}, {Batch}, {Baune}, {Bavigadda}, {Bazzan},
  {Bejger}, {Bell}, {Berger}, {Bergmann}, {Berry}, {Bersanetti}, {Bertolini},
  {Betzwieser}, {Bhagwat}, {Bhandare}, {Bilenko}, {Billingsley}, {Birch},
  {Birney}, {Birnholtz}, {Biscans}, {Bisht}, {Bitossi}, {Biwer}, {Bizouard},
  {Blackburn}, {Blair}, {Blair}, {Blair}, {Bloemen}, {Bock}, {Boer}, {Bogaert},
  {Bogan}, {Bohe}, {Bond}, {Bondu}, {Bonnand}, {Boom}, {Bork}, {Boschi},
  {Bose}, {Bouffanais}, {Bozzi}, {Bradaschia}, {Brady}, {Braginsky},
  {Branchesi}, {Brau}, {Briant}, {Brillet}, {Brinkmann}, {Brisson}, {Brockill},
  {Broida}, {Brooks}, {Brown}, {Brown}, {Brown}, {Brunett}, {Buchanan},
  {Buikema}, {Bulik}, {Bulten}, {Buonanno}, {Buskulic}, {Buy}, {Byer},
  {Cabero}, {Cadonati}, {Cagnoli}, {Cahillane}, {Calder{\'o}n Bustillo},
  {Callister}, {Calloni}, {Camp}, {Cannon}, {Cao}, {Capano}, {Capocasa},
  {Carbognani}, {Caride}, {Casanueva Diaz}, {Casentini}, {Caudill},
  {Cavagli{\`a}}, {Cavalier}, {Cavalieri}, {Cella}, {Cepeda}, {Cerboni
  Baiardi}, {Cerretani}, {Cesarini}, {Chamberlin}, {Chan}, {Chao}, {Charlton},
  {Chassande-Mottin}, {Cheeseboro}, {Chen}, {Chen}, {Cheng}, {Chincarini},
  {Chiummo}, {Cho}, {Cho}, {Chow}, {Christensen}, {Chu}, {Chua}, {Chung},
  {Ciani}, {Clara}, {Clark}, {Cleva}, {Coccia}, {Cohadon}, {Colla}, {Collette},
  {Cominsky}, {Constancio}, {Conte}, {Conti}, {Cook}, {Corbitt}, {Cornish},
  {Corsi}, {Cortese}, {Costa}, {Coughlin}, {Coughlin}, {Coulon}, {Countryman},
  {Couvares}, {Cowan}, {Coward}, {Cowart}, {Coyne}, {Coyne}, {Craig},
  {Creighton}, {Cripe}, {Crowder}, {Cumming}, {Cunningham}, {Cuoco}, {Dal
  Canton}, {Danilishin}, {D'Antonio}, {Danzmann}, {Darman}, {Dasgupta}, {Da
  Silva Costa}, {Dattilo}, {Dave}, {Davier}, {Davies}, {Daw}, {Day}, {De},
  {DeBra}, {Debreczeni}, {Degallaix}, {De Laurentis}, {Del{\'e}glise}, {Del
  Pozzo}, {Denker}, {Dent}, {Dergachev}, {De Rosa}, {DeRosa}, {DeSalvo},
  {Devine}, {Dhurandhar}, {D{\'\i}az}, {Di Fiore}, {Di Giovanni}, {Di
  Girolamo}, {Di Lieto}, {Di Pace}, {Di Palma}, {Di Virgilio}, {Dolique},
  {Donovan}, {Dooley}, {Doravari}, {Douglas}, {Downes}, {Drago}, {Drever},
  {Driggers}, {Ducrot}, {Dwyer}, {Edo}, {Edwards}, {Effler}, {Eggenstein},
  {Ehrens}, {Eichholz}, {Eikenberry}, {Engels}, {Essick}, {Etzel}, {Evans},
  {Evans}, {Everett}, {Factourovich}, {Fafone}, {Fair}, {Fairhurst}, {Fan},
  {Fang}, {Farinon}, {Farr}, {Farr}, {Favata}, {Fays}, {Fehrmann}, {Fejer},
  {Fenyvesi}, {Ferrante}, {Ferreira}, {Ferrini}, {Fidecaro}, {Fiori},
  {Fiorucci}, {Fisher}, {Flaminio}, {Fletcher}, {Fong}, {Fournier}, {Frasca},
  {Frasconi}, {Frei}, {Freise}, {Frey}, {Frey}, {Fritschel}, {Frolov}, {Fulda},
  {Fyffe}, {Gabbard}, {Gair}, {Gammaitoni}, {Gaonkar}, {Garufi}, {Gaur},
  {Gehrels}, {Gemme}, {Geng}, {Genin}, {Gennai}, {George}, {Gergely},
  {Germain}, {Ghosh}, {Ghosh}, {Ghosh}, {Giaime}, {Giardina}, {Giazotto},
  {Gill}, {Glaefke}, {Goetz}, {Goetz}, {Gondan}, {Gonz{\'a}lez}, {Gonzalez
  Castro}, {Gopakumar}, {Gordon}, {Gorodetsky}, {Gossan}, {Gosselin}, {Gouaty},
  {Grado}, {Graef}, {Graff}, {Granata}, {Grant}, {Gras}, {Gray}, {Greco},
  {Green}, {Groot}, {Grote}, {Grunewald}, {Guidi}, {Guo}, {Gupta}, {Gupta},
  {Gushwa}, {Gustafson}, {Gustafson}, {Hacker}, {Hall}, {Hall}, {Hamilton},
  {Hammond}, {Haney}, {Hanke}, {Hanks}, {Hanna}, {Hannam}, {Hanson},
  {Hardwick}, {Harms}, {Harry}, {Harry}, {Hart}, {Hartman}, {Haster},
  {Haughian}, {Healy}, {Heidmann}, {Heintze}, {Heitmann}, {Hello}, {Hemming},
  {Hendry}, {Heng}, {Hennig}, {Henry}, {Heptonstall}, {Heurs}, {Hild}, {Hoak},
  {Hofman}, {Holt}, {Holz}, {Hopkins}, {Hough}, {Houston}, {Howell}, {Hu},
  {Huang}, {Huerta}, {Huet}, {Hughey}, {Husa}, {Huttner}, {Huynh-Dinh},
  {Indik}, {Ingram}, {Inta}, {Isa}, {Isac}, {Isi}, {Isogai}, {Iyer}, {Izumi},
  {Jacqmin}, {Jang}, {Jani}, {Jaranowski}, {Jawahar}, {Jian},
  {Jim{\'e}nez-Forteza}, {Johnson}, {Johnson-McDaniel}, {Jones}, {Jones},
  {Jonker}, {Ju}, {K}, {Kalaghatgi}, {Kalogera}, {Kandhasamy}, {Kang},
  {Kanner}, {Kapadia}, {Karki}, {Karvinen}, {Kasprzack}, {Katsavounidis},
  {Katzman}, {Kaufer}, {Kaur}, {Kawabe}, {K{\'e}f{\'e}lian}, {Kehl}, {Keitel},
  {Kelley}, {Kells}, {Kennedy}, {Key}, {Khalili}, {Khan}, {Khan}, {Khan},
  {Khazanov}, {Kijbunchoo}, {Kim}, {Kim}, {Kim}, {Kim}, {Kim}, {Kim}, {Kim},
  {Kimbrell}, {King}, {King}, {Kissel}, {Klein}, {Kleybolte}, {Klimenko},
  {Koehlenbeck}, {Koley}, {Kondrashov}, {Kontos}, {Korobko}, {Korth},
  {Kowalska}, {Kozak}, {Kringel}, {Krishnan}, {Kr{\'o}lak}, {Krueger}, {Kuehn},
  {Kumar}, {Kumar}, {Kuo}, {Kutynia}, {Lackey}, {Landry}, {Lange}, {Lantz},
  {Lasky}, {Laxen}, {Lazzarini}, {Lazzaro}, {Leaci}, {Leavey}, {Lebigot},
  {Lee}, {Lee}, {Lee}, {Lee}, {Lenon}, {Leonardi}, {Leong}, {Leroy},
  {Letendre}, {Levin}, {Lewis}, {Li}, {Libson}, {Littenberg}, {Lockerbie},
  {Lombardi}, {London}, {Lord}, {Lorenzini}, {Loriette}, {Lormand}, {Losurdo},
  {Lough}, {Lousto}, {L{\"u}ck}, {Lundgren}, {Lynch}, {Ma}, {Machenschalk},
  {MacInnis}, {Macleod}, {Maga{\~n}a-Sandoval}, {Maga{\~n}a Zertuche}, {Magee},
  {Majorana}, {Maksimovic}, {Malvezzi}, {Man}, {Mandel}, {Mandic}, {Mangano},
  {Mansell}, {Manske}, {Mantovani}, {Marchesoni}, {Marion}, {M{\'a}rka},
  {M{\'a}rka}, {Markosyan}, {Maros}, {Martelli}, {Martellini}, {Martin},
  {Martynov}, {Marx}, {Mason}, {Masserot}, {Massinger}, {Masso-Reid},
  {Mastrogiovanni}, {Matichard}, {Matone}, {Mavalvala}, {Mazumder}, {McCarthy},
  {McClelland}, {McCormick}, {McGuire}, {McIntyre}, {McIver}, {McManus},
  {McRae}, {McWilliams}, {Meacher}, {Meadors}, {Meidam}, {Melatos}, {Mendell},
  {Mercer}, {Merilh}, {Merzougui}, {Meshkov}, {Messenger}, {Messick},
  {Metzdorff}, {Meyers}, {Mezzani}, {Miao}, {Michel}, {Middleton}, {Mikhailov},
  {Milano}, {Miller}, {Miller}, {Miller}, {Miller}, {Millhouse}, {Minenkov},
  {Ming}, {Mirshekari}, {Mishra}, {Mitra}, {Mitrofanov}, {Mitselmakher},
  {Mittleman}, {Moggi}, {Mohan}, {Mohapatra}, {Montani}, {Moore}, {Moore},
  {Moraru}, {Moreno}, {Morriss}, {Mossavi}, {Mours}, {Mow-Lowry}, {Mueller},
  {Muir}, {Mukherjee}, {Mukherjee}, {Mukherjee}, {Mukund}, {Mullavey}, {Munch},
  {Murphy}, {Murray}, {Mytidis}, {Nardecchia}, {Naticchioni}, {Nayak},
  {Nedkova}, {Nelemans}, {Nelson}, {Neri}, {Neunzert}, {Newton}, {Nguyen},
  {Nielsen}, {Nissanke}, {Nitz}, {Nocera}, {Nolting}, {Normandin}, {Nuttall},
  {Oberling}, {Ochsner}, {O'Dell}, {Oelker}, {Ogin}, {Oh}, {Oh}, {Ohme},
  {Oliver}, {Oppermann}, {Oram}, {O'Reilly}, {O'Shaughnessy}, {Ottaway},
  {Overmier}, {Owen}, {Pai}, {Pai}, {Palamos}, {Palashov}, {Palomba},
  {Pal-Singh}, {Pan}, {Pankow}, {Pannarale}, {Pant}, {Paoletti}, {Paoli},
  {Papa}, {Paris}, {Parker}, {Pascucci}, {Pasqualetti}, {Passaquieti},
  {Passuello}, {Patricelli}, {Patrick}, {Pearlstone}, {Pedraza}, {Pedurand},
  {Pekowsky}, {Pele}, {Penn}, {Perreca}, {Perri}, {Pfeiffer}, {Phelps},
  {Piccinni}, {Pichot}, {Piergiovanni}, {Pierro}, {Pillant}, {Pinard}, {Pinto},
  {Pitkin}, {Poe}, {Poggiani}, {Popolizio}, {Post}, {Powell}, {Prasad},
  {Predoi}, {Prestegard}, {Price}, {Prijatelj}, {Principe}, {Privitera},
  {Prix}, {Prodi}, {Prokhorov}, {Puncken}, {Punturo}, {Puppo}, {P{\"u}rrer},
  {Qi}, {Qin}, {Qiu}, {Quetschke}, {Quintero}, {Quitzow-James}, {Raab},
  {Rabeling}, {Radkins}, {Raffai}, {Raja}, {Rajan}, {Rakhmanov}, {Rapagnani},
  {Raymond}, {Razzano}, {Re}, {Read}, {Reed}, {Regimbau}, {Rei}, {Reid},
  {Reitze}, {Rew}, {Reyes}, {Ricci}, {Riles}, {Rizzo}, {Robertson}, {Robie},
  {Robinet}, {Rocchi}, {Rolland}, {Rollins}, {Roma}, {Romano}, {Romano},
  {Romanov}, {Romie}, {Rosi{\'n}ska}, {Rowan}, {R{\"u}diger}, {Ruggi}, {Ryan},
  {Sachdev}, {Sadecki}, {Sadeghian}, {Sakellariadou}, {Salconi}, {Saleem},
  {Salemi}, {Samajdar}, {Sammut}, {Sanchez}, {Sandberg}, {Sandeen}, {Sanders},
  {Sassolas}, {Sathyaprakash}, {Saulson}, {Sauter}, {Savage}, {Sawadsky},
  {Schale}, {Schilling}, {Schmidt}, {Schmidt}, {Schnabel}, {Schofield},
  {Sch{\"o}nbeck}, {Schreiber}, {Schuette}, {Schutz}, {Scott}, {Scott},
  {Sellers}, {Sengupta}, {Sentenac}, {Sequino}, {Sergeev}, {Setyawati},
  {Shaddock}, {Shaffer}, {Shahriar}, {Shaltev}, {Shapiro}, {Shawhan},
  {Sheperd}, {Shoemaker}, {Shoemaker}, {Siellez}, {Siemens}, {Sieniawska},
  {Sigg}, {Silva}, {Singer}, {Singer}, {Singh}, {Singh}, {Singhal}, {Sintes},
  {Slagmolen}, {Smith}, {Smith}, {Smith}, {Son}, {Sorazu}, {Sorrentino},
  {Souradeep}, {Srivastava}, {Staley}, {Steinke}, {Steinlechner},
  {Steinlechner}, {Steinmeyer}, {Stephens}, {Stevenson}, {Stone}, {Strain},
  {Straniero}, {Stratta}, {Strauss}, {Strigin}, {Sturani}, {Stuver},
  {Summerscales}, {Sun}, {Sunil}, {Sutton}, {Swinkels}, {Szczepa{\'n}czyk},
  {Tacca}, {Talukder}, {Tanner}, {T{\'a}pai}, {Tarabrin}, {Taracchini},
  {Taylor}, {Theeg}, {Thirugnanasambandam}, {Thomas}, {Thomas}, {Thomas},
  {Thorne}, {Thrane}, {Tiwari}, {Tiwari}, {Tokmakov}, {Toland}, {Tomlinson},
  {Tonelli}, {Tornasi}, {Torres}, {Torrie}, {T{\"o}yr{\"a}}, {Travasso},
  {Traylor}, {Trifir{\`o}}, {Tringali}, {Trozzo}, {Tse}, {Turconi},
  {Tuyenbayev}, {Ugolini}, {Unnikrishnan}, {Urban}, {Usman}, {Vahlbruch},
  {Vajente}, {Valdes}, {Vallisneri}, {van Bakel}, {van Beuzekom}, {van den
  Brand}, {Van Den Broeck}, {Vander-Hyde}, {van der Schaaf}, {van Heijningen},
  {van Veggel}, {Vardaro}, {Vass}, {Vas{\'u}th}, {Vaulin}, {Vecchio},
  {Vedovato}, {Veitch}, {Veitch}, {Venkateswara}, {Verkindt}, {Vetrano},
  {Vicer{\'e}}, {Vinciguerra}, {Vine}, {Vinet}, {Vitale}, {Vo}, {Vocca},
  {Vorvick}, {Voss}, {Vousden}, {Vyatchanin}, {Wade}, {Wade}, {Wade}, {Walker},
  {Wallace}, {Walsh}, {Wang}, {Wang}, {Wang}, {Wang}, {Wang}, {Ward}, {Warner},
  {Was}, {Weaver}, {Wei}, {Weinert}, {Weinstein}, {Weiss}, {Wen}, {We{\ss}els},
  {Westphal}, {Wette}, {Whelan}, {Whiting}, {Williams}, {Williamson}, {Willis},
  {Willke}, {Wimmer}, {Winkler}, {Wipf}, {Wittel}, {Woan}, {Woehler}, {Worden},
  {Wright}, {Wu}, {Wu}, {Yablon}, {Yam}, {Yamamoto}, {Yancey}, {Yu}, {Yvert},
  {Zadro{\.z}ny}, {Zangrando}, {Zanolin}, {Zendri}, {Zevin}, {Zhang}, {Zhang},
  {Zhang}, {Zhao}, {Zhou}, {Zhou}, {Zhu}, {Zucker}, {Zuraw}, {Zweizig},
  {Boyle}, {Hemberger}, {Kidder}, {Lovelace}, {Ossokine}, {Scheel}, {Szilagyi},
  {Teukolsky}, {LIGO Scientific Collaboration}, \& {VIRGO
  Collaboration}}]{2016PhRvL.116x1103A}
{Abbott}, B.~P., {Abbott}, R., {Abbott}, T.~D., {et~al.} 2016, \prl, 116,
  241103, \dodoi{10.1103/PhysRevLett.116.241103}

\bibitem[{{Abbott} {et~al.}(2019){Abbott}, {Abbott}, {Abbott}, {Abraham},
  {Acernese}, {Ackley}, {Adams}, {Adhikari}, {Adya}, {Affeldt}, {Agathos},
  {Agatsuma}, {Aggarwal}, {Aguiar}, {Aiello}, {Ain}, {Ajith}, {Allen},
  {Allocca}, {Aloy}, {Altin}, {Amato}, {Anand}, {Ananyeva}, {Anderson},
  {Anderson}, {Angelova}, {Antier}, {Appert}, {Arai}, {Araya}, {Areeda},
  {Ar{\`e}ne}, {Arnaud}, {Aronson}, {Ascenzi}, {Ashton}, {Aston}, {Astone},
  {Aubin}, {Aufmuth}, {AultONeal}, {Austin}, {Avendano}, {Avila-Alvarez},
  {Babak}, {Bacon}, {Badaracco}, {Bader}, {Bae}, {Baird}, {Baker},
  {Baldaccini}, {Ballardin}, {Ballmer}, {Bals}, {Banagiri}, {Barayoga},
  {Barbieri}, {Barclay}, {Barish}, {Barker}, {Barkett}, {Barnum}, {Barone},
  {Barr}, {Barsotti}, {Barsuglia}, {Barta}, {Bartlett}, {Bartos}, {Bassiri},
  {Basti}, {Bawaj}, {Bayley}, {Bazzan}, {B{\'e}csy}, {Bejger}, {Belahcene},
  {Bell}, {Beniwal}, {Benjamin}, {Berger}, {Bergmann}, {Bernuzzi}, {Berry},
  {Bersanetti}, {Bertolini}, {Betzwieser}, {Bhandare}, {Bidler}, {Biggs},
  {Bilenko}, {Bilgili}, {Billingsley}, {Birney}, {Birnholtz}, {Biscans},
  {Bischi}, {Biscoveanu}, {Bisht}, {Bitossi}, {Bizouard}, {Blackburn},
  {Blackman}, {Blair}, {Blair}, {Blair}, {Bloemen}, {Bobba}, {Bode}, {Boer},
  {Boetzel}, {Bogaert}, {Bondu}, {Bonnand}, {Booker}, {Boom}, {Bork}, {Boschi},
  {Bose}, {Bossilkov}, {Bosveld}, {Bouffanais}, {Bozzi}, {Bradaschia}, {Brady},
  {Bramley}, {Branchesi}, {Brau}, {Breschi}, {Briant}, {Briggs}, {Brighenti},
  {Brillet}, {Brinkmann}, {Brockill}, {Brooks}, {Brooks}, {Brown}, {Brunett},
  {Buikema}, {Bulik}, {Bulten}, {Buonanno}, {Buskulic}, {Buy}, {Byer},
  {Cabero}, {Cadonati}, {Cagnoli}, {Cahillane}, {Calder{\'o}n Bustillo},
  {Callister}, {Calloni}, {Camp}, {Campbell}, {Canepa}, {Cannon}, {Cao}, {Cao},
  {Carapella}, {Carbognani}, {Caride}, {Carney}, {Carullo}, {Casanueva Diaz},
  {Casentini}, {Caudill}, {Cavagli{\`a}}, {Cavalier}, {Cavalieri}, {Cella},
  {Cerd{\'a}-Dur{\'a}n}, {Cesarini}, {Chaibi}, {Chakravarti}, {Chamberlin},
  {Chan}, {Chao}, {Charlton}, {Chase}, {Chassande-Mottin}, {Chatterjee},
  {Chaturvedi}, {Chatziioannou}, {Cheeseboro}, {Chen}, {Chen}, {Chen}, {Cheng},
  {Cheong}, {Chia}, {Chiadini}, {Chincarini}, {Chiummo}, {Cho}, {Cho}, {Cho},
  {Christensen}, {Chu}, {Chua}, {Chung}, {Chung}, {Ciani}, {Cie{\'s}lar},
  {Ciobanu}, {Ciolfi}, {Cipriano}, {Cirone}, {Clara}, {Clark}, {Clearwater},
  {Cleva}, {Coccia}, {Cohadon}, {Cohen}, {Colleoni}, {Collette}, {Collins},
  {Colpi}, {Cominsky}, {Constancio}, {Conti}, {Cooper}, {Corban}, {Corbitt},
  {Cordero-Carri{\'o}n}, {Corezzi}, {Corley}, {Cornish}, {Corre}, {Corsi},
  {Cortese}, {Costa}, {Cotesta}, {Coughlin}, {Coughlin}, {Coulon},
  {Countryman}, {Couvares}, {Covas}, {Cowan}, {Coward}, {Cowart}, {Coyne},
  {Coyne}, {Creighton}, {Creighton}, {Cripe}, {Croquette}, {Crowder}, {Cullen},
  {Cumming}, {Cunningham}, {Cuoco}, {Dal Canton}, {D{\'a}lya}, {D'Angelo},
  {Danilishin}, {D'Antonio}, {Danzmann}, {Dasgupta}, {Da Silva Costa},
  {Datrier}, {Dattilo}, {Dave}, {Davier}, {Davis}, {Daw}, {DeBra},
  {Deenadayalan}, {Degallaix}, {De Laurentis}, {Del{\'e}glise}, {Del Pozzo},
  {DeMarchi}, {Demos}, {Dent}, {De Pietri}, {De Rosa}, {De Rossi}, {DeSalvo},
  {de Varona}, {Dhurandhar}, {D{\'\i}az}, {Dietrich}, {Di Fiore}, {DiFronzo},
  {Di Giorgio}, {Di Giovanni}, {Di Giovanni}, {Di Girolamo}, {Di Lieto},
  {Ding}, {Di Pace}, {Di Palma}, {Di Renzo}, {Divakarla}, {Dmitriev}, {Doctor},
  {Donovan}, {Dooley}, {Doravari}, {Dorrington}, {Downes}, {Drago}, {Driggers},
  {Du}, {Ducoin}, {Dupej}, {Durante}, {Dwyer}, {Easter}, {Eddolls}, {Edo},
  {Effler}, {Ehrens}, {Eichholz}, {Eikenberry}, {Eisenmann}, {Eisenstein},
  {Errico}, {Essick}, {Estelles}, {Estevez}, {Etienne}, {Etzel}, {Evans},
  {Evans}, {Fafone}, {Fairhurst}, {Fan}, {Farinon}, {Farr}, {Farr},
  {Fauchon-Jones}, {Favata}, {Fays}, {Fazio}, {Fee}, {Feicht}, {Fejer}, {Feng},
  {Fernandez-Galiana}, {Ferrante}, {Ferreira}, {Ferreira}, {Fidecaro}, {Fiori},
  {Fiorucci}, {Fishbach}, {Fisher}, {Fishner}, {Fittipaldi}, {Fitz-Axen},
  {Fiumara}, {Flaminio}, {Fletcher}, {Floden}, {Flynn}, {Fong}, {Font},
  {Forsyth}, {Fournier}, {Vivanco}, {Frasca}, {Frasconi}, {Frei}, {Freise},
  {Frey}, {Frey}, {Fritschel}, {Frolov}, {Fronz{\`e}}, {Fulda}, {Fyffe},
  {Gabbard}, {Gadre}, {Gaebel}, {Gair}, {Gammaitoni}, {Gaonkar},
  {Garc{\'\i}a-Quir{\'o}s}, {Garufi}, {Gateley}, {Gaudio}, {Gaur}, {Gayathri},
  {Gemme}, {Genin}, {Gennai}, {George}, {George}, {Gergely}, {Ghonge}, {Ghosh},
  {Ghosh}, {Ghosh}, {Giacomazzo}, {Giaime}, {Giardina}, {Gibson}, {Gill},
  {Glover}, {Gniesmer}, {Godwin}, {Goetz}, {Goetz}, {Goncharov},
  {Gonz{\'a}lez}, {Gonzalez Castro}, {Gopakumar}, {Gossan}, {Gosselin},
  {Gouaty}, {Grace}, {Grado}, {Granata}, {Grant}, {Gras}, {Grassia}, {Gray},
  {Gray}, {Greco}, {Green}, {Green}, {Gretarsson}, {Grimaldi}, {Grimm},
  {Groot}, {Grote}, {Grunewald}, {Gruning}, {Guidi}, {Gulati}, {Guo}, {Gupta},
  {Gupta}, {Gupta}, {Gustafson}, {Gustafson}, {Haegel}, {Halim}, {Hall},
  {Hall}, {Hamilton}, {Hammond}, {Haney}, {Hanke}, {Hanks}, {Hanna}, {Hannam},
  {Hannuksela}, {Hansen}, {Hanson}, {Harder}, {Hardwick}, {Haris}, {Harms},
  {Harry}, {Harry}, {Hasskew}, {Haster}, {Haughian}, {Hayes}, {Healy},
  {Heidmann}, {Heintze}, {Heitmann}, {Hellman}, {Hello}, {Hemming}, {Hendry},
  {Heng}, {Hennig}, {Heurs}, {Hild}, {Hinderer}, {Hochheim}, {Hofman},
  {Holgado}, {Holland}, {Holt}, {Holz}, {Hopkins}, {Horst}, {Hough}, {Howell},
  {Hoy}, {Huang}, {H{\"u}bner}, {Huerta}, {Huet}, {Hughey}, {Hui}, {Husa},
  {Huttner}, {Huynh-Dinh}, {Idzkowski}, {Iess}, {Inchauspe}, {Ingram}, {Inta},
  {Intini}, {Irwin}, {Isa}, {Isac}, {Isi}, {Iyer}, {Jacqmin}, {Jadhav}, {Jani},
  {Janthalur}, {Jaranowski}, {Jariwala}, {Jenkins}, {Jiang}, {Johnson},
  {Jones}, {Jones}, {Jones}, {Jones}, {Jonker}, {Ju}, {Junker}, {Kalaghatgi},
  {Kalogera}, {Kamai}, {Kandhasamy}, {Kang}, {Kanner}, {Kapadia}, {Karki},
  {Kashyap}, {Kasprzack}, {Katsanevas}, {Katsavounidis}, {Katzman}, {Kaufer},
  {Kawabe}, {Keerthana}, {K{\'e}f{\'e}lian}, {Keitel}, {Kennedy}, {Key},
  {Khalili}, {Khan}, {Khan}, {Khazanov}, {Khetan}, {Khursheed}, {Kijbunchoo},
  {Kim}, {Kim}, {Kim}, {Kim}, {Kim}, {Kim}, {Kimball}, {King}, {Kinley-Hanlon},
  {Kirchhoff}, {Kissel}, {Kleybolte}, {Klika}, {Klimenko}, {Knowles}, {Koch},
  {Koehlenbeck}, {Koekoek}, {Koley}, {Kondrashov}, {Kontos}, {Koper},
  {Korobko}, {Korth}, {Kovalam}, {Kozak}, {Kr{\"a}mer}, {Kringel},
  {Krishnendu}, {Kr{\'o}lak}, {Krupinski}, {Kuehn}, {Kumar}, {Kumar}, {Kumar},
  {Kumar}, {Kuo}, {Kutynia}, {Kwang}, {Lackey}, {Laghi}, {Lai}, {Lam},
  {Landry}, {Lane}, {Lang}, {Lange}, {Lantz}, {Lanza}, {Lartaux-Vollard},
  {Lasky}, {Laxen}, {Lazzarini}, {Lazzaro}, {Leaci}, {Leavey}, {Lecoeuche},
  {Lee}, {Lee}, {Lee}, {Lee}, {Lee}, {Lee}, {Lehmann}, {Lenon}, {Leroy},
  {Letendre}, {Levin}, {Li}, {Li}, {Li}, {Li}, {Li}, {Lin}, {Linde}, {Linker},
  {Littenberg}, {Liu}, {Liu}, {Llorens-Monteagudo}, {Lo}, {London}, {Longo},
  {Lorenzini}, {Loriette}, {Lormand}, {Losurdo}, {Lough}, {Lousto}, {Lovelace},
  {Lower}, {L{\"u}ck}, {Lumaca}, {Lundgren}, {Lynch}, {Ma}, {Macas}, {Macfoy},
  {MacInnis}, {Macleod}, {Macquet}, {Maga{\~n}a Hernandez},
  {Maga{\~n}a-Sandoval}, {Magee}, {Majorana}, {Maksimovic}, {Malik}, {Man},
  {Mandic}, {Mangano}, {Mansell}, {Manske}, {Mantovani}, {Mapelli},
  {Marchesoni}, {Marion}, {M{\'a}rka}, {M{\'a}rka}, {Markakis}, {Markosyan},
  {Markowitz}, {Maros}, {Marquina}, {Marsat}, {Martelli}, {Martin}, {Martin},
  {Martinez}, {Martynov}, {Masalehdan}, {Mason}, {Massera}, {Masserot},
  {Massinger}, {Masso-Reid}, {Mastrogiovanni}, {Matas}, {Matichard}, {Matone},
  {Mavalvala}, {McCann}, {McCarthy}, {McClelland}, {McCormick}, {McCuller},
  {McGuire}, {McIsaac}, {McIver}, {McManus}, {McRae}, {McWilliams}, {Meacher},
  {Meadors}, {Mehmet}, {Mehta}, {Meidam}, {Mejuto Villa}, {Melatos}, {Mendell},
  {Mercer}, {Mereni}, {Merfeld}, {Merilh}, {Merzougui}, {Meshkov}, {Messenger},
  {Messick}, {Messina}, {Metzdorff}, {Meyers}, {Meylahn}, {Miani}, {Miao},
  {Michel}, {Middleton}, {Milano}, {Miller}, {Millhouse}, {Mills},
  {Milovich-Goff}, {Minazzoli}, {Minenkov}, {Mishkin}, {Mishra}, {Mistry},
  {Mitra}, {Mitrofanov}, {Mitselmakher}, {Mittleman}, {Mo}, {Moffa}, {Mogushi},
  {Mohapatra}, {Molina-Ruiz}, {Mondin}, {Montani}, {Moore}, {Moraru},
  {Morawski}, {Moreno}, {Morisaki}, {Mours}, {Mow-Lowry}, {Muciaccia},
  {Mukherjee}, {Mukherjee}, {Mukherjee}, {Mukherjee}, {Mukund}, {Mullavey},
  {Munch}, {Mu{\~n}iz}, {Muratore}, {Murray}, {Nagar}, {Nardecchia},
  {Naticchioni}, {Nayak}, {Neil}, {Neilson}, {Nelemans}, {Nelson}, {Nery},
  {Neunzert}, {Nevin}, {Ng}, {Ng}, {Nguyen}, {Nguyen}, {Nichols}, {Nichols},
  {Nissanke}, {Nocera}, {North}, {Nuttall}, {Obergaulinger}, {Oberling},
  {O'Brien}, {Oganesyan}, {Ogin}, {Oh}, {Oh}, {Ohme}, {Ohta}, {Okada},
  {Oliver}, {Oppermann}, {Oram}, {O'Reilly}, {Ormiston}, {Ortega},
  {O'Shaughnessy}, {Ossokine}, {Ottaway}, {Overmier}, {Owen}, {Pace}, {Pagano},
  {Page}, {Pagliaroli}, {Pai}, {Pai}, {Palamos}, {Palashov}, {Palomba}, {Pan},
  {Panda}, {Pang}, {Pankow}, {Pannarale}, {Pant}, {Paoletti}, {Paoli},
  {Parida}, {Parker}, {Pascucci}, {Pasqualetti}, {Passaquieti}, {Passuello},
  {Patil}, {Patricelli}, {Payne}, {Pearlstone}, {Pechsiri}, {Pedersen},
  {Pedraza}, {Pedurand}, {Pele}, {Penn}, {Perego}, {Perez}, {P{\'e}rigois},
  {Perreca}, {Petermann}, {Pfeiffer}, {Phelps}, {Phukon}, {Piccinni}, {Pichot},
  {Piergiovanni}, {Pierro}, {Pillant}, {Pinard}, {Pinto}, {Pirello}, {Pitkin},
  {Plastino}, {Poggiani}, {Pong}, {Ponrathnam}, {Popolizio}, {Porter},
  {Powell}, {Prajapati}, {Prasad}, {Prasai}, {Prasanna}, {Pratten},
  {Prestegard}, {Principe}, {Prodi}, {Prokhorov}, {Punturo}, {Puppo},
  {P{\"u}rrer}, {Qi}, {Quetschke}, {Quinonez}, {Raab}, {Raaijmakers},
  {Radkins}, {Radulesco}, {Raffai}, {Raja}, {Rajan}, {Rajbhandari},
  {Rakhmanov}, {Ramirez}, {Ramos-Buades}, {Rana}, {Rao}, {Rapagnani},
  {Raymond}, {Razzano}, {Read}, {Regimbau}, {Rei}, {Reid}, {Reitze},
  {Rettegno}, {Ricci}, {Richardson}, {Richardson}, {Ricker}, {Riemenschneider},
  {Riles}, {Rizzo}, {Robertson}, {Robinet}, {Rocchi}, {Rolland}, {Rollins},
  {Roma}, {Romanelli}, {Romano}, {Romel}, {Romie}, {Rose}, {Rose}, {Rose},
  {Rosi{\'n}ska}, {Rosofsky}, {Ross}, {Rowan}, {R{\"u}diger}, {Ruggi},
  {Rutins}, {Ryan}, {Sachdev}, {Sadecki}, {Sakellariadou}, {Salafia},
  {Salconi}, {Saleem}, {Samajdar}, {Sammut}, {Sanchez}, {Sanchez},
  {Sanchis-Gual}, {Sanders}, {Santiago}, {Santos}, {Sarin}, {Sassolas},
  {Sauter}, {Savage}, {Schale}, {Scheel}, {Scheuer}, {Schmidt}, {Schnabel},
  {Schofield}, {Sch{\"o}nbeck}, {Schreiber}, {Schulte}, {Schutz}, {Scott},
  {Scott}, {Seidel}, {Sellers}, {Sengupta}, {Sennett}, {Sentenac}, {Sequino},
  {Sergeev}, {Setyawati}, {Shaddock}, {Shaffer}, {Shahriar}, {Shaner},
  {Sharma}, {Sharma}, {Shawhan}, {Shen}, {Shink}, {Shoemaker}, {Shoemaker},
  {Shukla}, {ShyamSundar}, {Siellez}, {Sieniawska}, {Sigg}, {Singer}, {Singh},
  {Singh}, {Singhal}, {Sintes}, {Sitmukhambetov}, {Skliris}, {Slagmolen},
  {Slaven-Blair}, {Smith}, {Smith}, {Somala}, {Son}, {Soni}, {Sorazu},
  {Sorrentino}, {Souradeep}, {Sowell}, {Spencer}, {Spera}, {Srivastava},
  {Srivastava}, {Staats}, {Stachie}, {Standke}, {Steer}, {Steinke},
  {Steinlechner}, {Steinlechner}, {Steinmeyer}, {Stevenson}, {Stocks}, {Stone},
  {Stops}, {Strain}, {Stratta}, {Strigin}, {Strunk}, {Sturani}, {Stuver},
  {Sudhir}, {Summerscales}, {Sun}, {Sunil}, {Sur}, {Suresh}, {Sutton},
  {Swinkels}, {Szczepa{\'n}czyk}, {Tacca}, {Tait}, {Talbot}, {Tanner}, {Tao},
  {T{\'a}pai}, {Tapia}, {Tasson}, {Taylor}, {Tenorio}, {Terkowski}, {Thomas},
  {Thomas}, {Thondapu}, {Thorne}, {Thrane}, {Tiwari}, {Tiwari}, {Tiwari},
  {Toland}, {Tonelli}, {Tornasi}, {Torres-Forn{\'e}}, {Torrie},
  {T{\"o}yr{\"a}}, {Travasso}, {Traylor}, {Tringali}, {Tripathee}, {Trovato},
  {Trozzo}, {Tsang}, {Tse}, {Tso}, {Tsukada}, {Tsuna}, {Tsutsui}, {Tuyenbayev},
  {Ueno}, {Ugolini}, {Unnikrishnan}, {Urban}, {Usman}, {Vahlbruch}, {Vajente},
  {Valdes}, {Valentini}, {van Bakel}, {van Beuzekom}, {van den Brand}, {Van Den
  Broeck}, {Vander-Hyde}, {van der Schaaf}, {VanHeijningen}, {van Veggel},
  {Vardaro}, {Varma}, {Vass}, {Vas{\'u}th}, {Vecchio}, {Vedovato}, {Veitch},
  {Veitch}, {Venkateswara}, {Venugopalan}, {Verkindt}, {Vetrano}, {Vicer{\'e}},
  {Viets}, {Vinciguerra}, {Vine}, {Vinet}, {Vitale}, {Vo}, {Vocca}, {Vorvick},
  {Vyatchanin}, {Wade}, {Wade}, {Wade}, {Walet}, {Walker}, {Wallace}, {Walsh},
  {Wang}, {Wang}, {Wang}, {Wang}, {Wang}, {Ward}, {Warden}, {Warner}, {Was},
  {Watchi}, {Weaver}, {Wei}, {Weinert}, {Weinstein}, {Weiss}, {Wellmann},
  {Wen}, {Wessel}, {We{\ss}els}, {Westhouse}, {Wette}, {Whelan}, {Whiting},
  {Whittle}, {Wilken}, {Williams}, {Williamson}, {Willis}, {Willke}, {Winkler},
  {Wipf}, {Wittel}, {Woan}, {Woehler}, {Wofford}, {Wright}, {Wu}, {Wysocki},
  {Xiao}, {Xu}, {Yamamoto}, {Yancey}, {Yang}, {Yang}, {Yang}, {Yap}, {Yazback},
  {Yeeles}, {Yu}, {Yu}, {Yuen}, {Zadro{\.Z}ny}, {Zadro{\.Z}ny}, {Zanolin},
  {Zelenova}, {Zendri}, {Zevin}, {Zhang}, {Zhang}, {Zhang}, {Zhao}, {Zhao},
  {Zhou}, {Zhou}, {Zhu}, {Zucker}, {Zweizig}, {Papa}, {Salemi}, {LIGO
  Scientific Collaboration}, \& {Virgo Collaboration}}]{2019PhRvD.100b4017A}
---. 2019, \prd, 100, 024017, \dodoi{10.1103/PhysRevD.100.024017}

\bibitem[{{Afshordi} {et~al.}(2003){Afshordi}, {McDonald}, \&
  {Spergel}}]{2003ApJ...594L..71A}
{Afshordi}, N., {McDonald}, P., \& {Spergel}, D.~N. 2003, \apjl, 594, L71,
  \dodoi{10.1086/378763}

\bibitem[{Afshordi {et~al.}(2003)Afshordi, McDonald, \&
  Spergel}]{Afshordi:2003zb}
Afshordi, N., McDonald, P., \& Spergel, D.~N. 2003, Astrophys. J., 594, L71,
  \dodoi{10.1086/378763}

\bibitem[{{Alcock} {et~al.}(2001){Alcock}, {Allsman}, {Alves}, {Axelrod},
  {Becker}, {Bennett}, {Cook}, {Dalal}, {Drake}, {Freeman}, {Geha}, {Griest},
  {Lehner}, {Marshall}, {Minniti}, {Nelson}, {Peterson}, {Popowski}, {Pratt},
  {Quinn}, {Stubbs}, {Sutherland}, {Tomaney}, {Vandehei}, \&
  {Welch}}]{2001ApJ...550L.169A}
{Alcock}, C., {Allsman}, R.~A., {Alves}, D.~R., {et~al.} 2001, \apjl, 550,
  L169, \dodoi{10.1086/319636}

\bibitem[{{Ali-Ha{\"\i}moud}(2018)}]{2018PhRvL.121h1304A}
{Ali-Ha{\"\i}moud}, Y. 2018, \prl, 121, 081304,
  \dodoi{10.1103/PhysRevLett.121.081304}

\bibitem[{Ali-Ha{\"{\i}}moud \& Kamionkowski(2017)}]{Ali-Haimoud:2016mbv}
Ali-Ha{\"{\i}}moud, Y., \& Kamionkowski, M. 2017, Phys. Rev., D95, 043534,
  \dodoi{10.1103/PhysRevD.95.043534}

\bibitem[{{Ba{\~n}ados} {et~al.}(2018){Ba{\~n}ados}, {Carilli}, {Walter},
  {Momjian}, {Decarli}, {Farina}, {Mazzucchelli}, \& {Venemans}}]{ban2018}
{Ba{\~n}ados}, E., {Carilli}, C., {Walter}, F., {et~al.} 2018, ApJL, 861, L14,
  \dodoi{10.3847/2041-8213/aac511}

\bibitem[{Ba{\~n}ados {et~al.}(2015)}]{ban2015}
Ba{\~n}ados, E., {et~al.} 2015, Astrophys. J., 804, 118,
  \dodoi{10.1088/0004-637X/804/2/118}

\bibitem[{Ba{\~n}ados {et~al.}(2021)}]{Banados:2021imw}
---. 2021, Astrophys. J., 909, 80, \dodoi{10.3847/1538-4357/abe239}

\bibitem[{Barkana \& Loeb(2001)}]{Barkana:2000fd}
Barkana, R., \& Loeb, A. 2001, Phys. Rept., 349, 125,
  \dodoi{10.1016/S0370-1573(01)00019-9}

\bibitem[{Belladitta {et~al.}(2020)}]{Belladitta:2020zbp}
Belladitta, S., {et~al.} 2020, Astron. Astrophys., 635, L7,
  \dodoi{10.1051/0004-6361/201937395}

\bibitem[{Boehm {et~al.}(2021)Boehm, Kobakhidze, O'hare, Picker, \&
  Sakellariadou}]{Boehm:2020jwd}
Boehm, C., Kobakhidze, A., O'hare, C. A.~J., Picker, Z. S.~C., \&
  Sakellariadou, M. 2021, JCAP, 03, 078, \dodoi{10.1088/1475-7516/2021/03/078}

\bibitem[{{Bondi}(1952)}]{1952MNRAS.112..195B}
{Bondi}, H. 1952, \mnras, 112, 195, \dodoi{10.1093/mnras/112.2.195}

\bibitem[{Bondi \& Hoyle(1944)}]{bondi1944mechanism}
Bondi, H., \& Hoyle, F. 1944, Mon. Not. Roy. Astron. Soc., 104, 273

\bibitem[{{Bowman} {et~al.}(2018){Bowman}, {Rogers}, {Monsalve}, {Mozdzen}, \&
  {Mahesh}}]{2018Natur.555...67B}
{Bowman}, J.~D., {Rogers}, A. E.~E., {Monsalve}, R.~A., {Mozdzen}, T.~J., \&
  {Mahesh}, N. 2018, \nat, 555, 67, \dodoi{10.1038/nature25792}

\bibitem[{{Brandt}(2016)}]{2016ApJ...824L..31B}
{Brandt}, T.~D. 2016, \apjl, 824, L31, \dodoi{10.3847/2041-8205/824/2/L31}

\bibitem[{Bryan \& Norman(1998)}]{Bryan:1997dn}
Bryan, G.~L., \& Norman, M.~L. 1998, Astrophys. J., 495, 80,
  \dodoi{10.1086/305262}

\bibitem[{{Bullock} {et~al.}(2001){Bullock}, {Kolatt}, {Sigad}, {Somerville},
  {Kravtsov}, {Klypin}, {Primack}, \& {Dekel}}]{2001MNRAS.321..559B}
{Bullock}, J.~S., {Kolatt}, T.~S., {Sigad}, Y., {et~al.} 2001, \mnras, 321,
  559, \dodoi{10.1046/j.1365-8711.2001.04068.x}

\bibitem[{{Carilli} {et~al.}(2002){Carilli}, {Gnedin}, \&
  {Owen}}]{2002ApJ...577...22C}
{Carilli}, C.~L., {Gnedin}, N.~Y., \& {Owen}, F. 2002, \apj, 577, 22,
  \dodoi{10.1086/342179}

\bibitem[{{Carr} {et~al.}(2020){Carr}, {Kohri}, {Sendouda}, \&
  {Yokoyama}}]{2020arXiv200212778C}
{Carr}, B., {Kohri}, K., {Sendouda}, Y., \& {Yokoyama}, J. 2020, arXiv
  e-prints, arXiv:2002.12778.
\newblock \doarXiv{2002.12778}

\bibitem[{{Carr} {et~al.}(2016){Carr}, {K{\"u}hnel}, \&
  {Sandstad}}]{2016PhRvD..94h3504C}
{Carr}, B., {K{\"u}hnel}, F., \& {Sandstad}, M. 2016, \prd, 94, 083504,
  \dodoi{10.1103/PhysRevD.94.083504}

\bibitem[{Carr \& Silk(2018)}]{Carr:2018rid}
Carr, B., \& Silk, J. 2018, Mon. Not. Roy. Astron. Soc., 478, 3756,
  \dodoi{10.1093/mnras/sty1204}

\bibitem[{{Chen} \& {Huang}(2020)}]{2020JCAP...08..039C}
{Chen}, Z.-C., \& {Huang}, Q.-G. 2020, \jcap, 2020, 039,
  \dodoi{10.1088/1475-7516/2020/08/039}

\bibitem[{Child {et~al.}(2018)Child, Habib, Heitmann, Frontiere, Finkel, Pope,
  \& Morozov}]{Child:2018skq}
Child, H.~L., Habib, S., Heitmann, K., {et~al.} 2018, Astrophys. J., 859, 55,
  \dodoi{10.3847/1538-4357/aabf95}

\bibitem[{{Chisholm}(2006)}]{2006PhRvD..73h3504C}
{Chisholm}, J.~R. 2006, \prd, 73, 083504, \dodoi{10.1103/PhysRevD.73.083504}

\bibitem[{{Chluba} \& {Thomas}(2011)}]{2011MNRAS.412..748C}
{Chluba}, J., \& {Thomas}, R.~M. 2011, \mnras, 412, 748,
  \dodoi{10.1111/j.1365-2966.2010.17940.x}

\bibitem[{{Ciardi} {et~al.}(2013){Ciardi}, {Labropoulos}, {Maselli}, {Thomas},
  {Zaroubi}, {Graziani}, {Bolton}, {Bernardi}, {Brentjens}, {de Bruyn},
  {Daiboo}, {Harker}, {Jelic}, {Kazemi}, {Koopmans}, {Martinez}, {Mellema},
  {Offringa}, {Pandey}, {Schaye}, {Veligatla}, {Vedantham}, \&
  {Yatawatta}}]{2013MNRAS.428.1755C}
{Ciardi}, B., {Labropoulos}, P., {Maselli}, A., {et~al.} 2013, \mnras, 428,
  1755, \dodoi{10.1093/mnras/sts156}

\bibitem[{Ciardi {et~al.}(2015)}]{Ciardi:2015lia}
Ciardi, B., {et~al.} 2015, Mon. Not. Roy. Astron. Soc., 453, 101,
  \dodoi{10.1093/mnras/stv1640}

\bibitem[{{Clesse} \& {Garc{\'\i}a-Bellido}(2017)}]{2017PDU....15..142C}
{Clesse}, S., \& {Garc{\'\i}a-Bellido}, J. 2017, Physics of the Dark Universe,
  15, 142, \dodoi{10.1016/j.dark.2016.10.002}

\bibitem[{{Cucchiara} {et~al.}(2011){Cucchiara}, {Levan}, {Fox}, {Tanvir},
  {Ukwatta}, {Berger}, {Kr{\"u}hler}, {K{\"u}pc{\"u} Yoldas}, {Wu}, {Toma},
  {Greiner}, {Olivares}, {Rowlinson}, {Amati}, {Sakamoto}, {Roth}, {Stephens},
  {Fritz}, {Fynbo}, {Hjorth}, {Malesani}, {Jakobsson}, {Wiersema}, {O'Brien},
  {Soderberg}, {Foley}, {Fruchter}, {Rhoads}, {Rutledge}, {Schmidt}, {Dopita},
  {Podsiadlowski}, {Willingale}, {Wolf}, {Kulkarni}, \&
  {D'Avanzo}}]{2011ApJ...736....7C}
{Cucchiara}, A., {Levan}, A.~J., {Fox}, D.~B., {et~al.} 2011, \apj, 736, 7,
  \dodoi{10.1088/0004-637X/736/1/7}

\bibitem[{{de Souza} {et~al.}(2011){de Souza}, {Yoshida}, \&
  {Ioka}}]{2011A&A...533A..32D}
{de Souza}, R.~S., {Yoshida}, N., \& {Ioka}, K. 2011, \aap, 533, A32,
  \dodoi{10.1051/0004-6361/201117242}

\bibitem[{{Desjacques} \& {Riotto}(2018)}]{2018PhRvD..98l3533D}
{Desjacques}, V., \& {Riotto}, A. 2018, \prd, 98, 123533,
  \dodoi{10.1103/PhysRevD.98.123533}

\bibitem[{{Diemer}(2018)}]{2018ApJS..239...35D}
{Diemer}, B. 2018, \apjs, 239, 35, \dodoi{10.3847/1538-4365/aaee8c}

\bibitem[{Diemer \& Joyce(2019)}]{Diemer:2018vmz}
Diemer, B., \& Joyce, M. 2019, Astrophys. J., 871, 168,
  \dodoi{10.3847/1538-4357/aafad6}

\bibitem[{Furlanetto(2006)}]{Furlanetto:2006dt}
Furlanetto, S. 2006, Mon. Not. Roy. Astron. Soc., 370, 1867,
  \dodoi{10.1111/j.1365-2966.2006.10603.x}

\bibitem[{Furlanetto {et~al.}(2006)Furlanetto, Oh, \&
  Briggs}]{Furlanetto:2006jb}
Furlanetto, S., Oh, S.~P., \& Briggs, F. 2006, Phys. Rept., 433, 181,
  \dodoi{10.1016/j.physrep.2006.08.002}

\bibitem[{{Furlanetto} \& {Loeb}(2002)}]{2002ApJ...579....1F}
{Furlanetto}, S.~R., \& {Loeb}, A. 2002, \apj, 579, 1, \dodoi{10.1086/342757}

\bibitem[{Gong \& Kitajima(2017)}]{Gong:2017sie}
Gong, J.-O., \& Kitajima, N. 2017, JCAP, 1708, 017,
  \dodoi{10.1088/1475-7516/2017/08/017}

\bibitem[{Gong \& Kitajima(2018)}]{Gong:2018sos}
---. 2018, JCAP, 1811, 041, \dodoi{10.1088/1475-7516/2018/11/041}

\bibitem[{{Green} \& {Kavanagh}(2020)}]{2020arXiv200710722G}
{Green}, A.~M., \& {Kavanagh}, B.~J. 2020, arXiv e-prints, arXiv:2007.10722.
\newblock \doarXiv{2007.10722}

\bibitem[{Greiner {et~al.}(2009)}]{Greiner:2008fb}
Greiner, J., {et~al.} 2009, Astrophys. J., 693, 1610,
  \dodoi{10.1088/0004-637X/693/2/1610}

\bibitem[{Griest {et~al.}(2014)Griest, Cieplak, \& Lehner}]{Griest:2013aaa}
Griest, K., Cieplak, A.~M., \& Lehner, M.~J. 2014, Astrophys. J., 786, 158,
  \dodoi{10.1088/0004-637X/786/2/158}

\bibitem[{Haiman {et~al.}(2004)Haiman, Quataert, \& Bower}]{Haiman_2004}
Haiman, Z., Quataert, E., \& Bower, G.~C. 2004, The Astrophysical Journal, 612,
  698, \dodoi{10.1086/422834}

\bibitem[{Ichimaru(1977)}]{Ichimaru:1977uf}
Ichimaru, S. 1977, Astrophys. J., 214, 840, \dodoi{10.1086/155314}

\bibitem[{Inman \& Ali-Ha\"\i{}moud(2019)}]{Inman:2019wvr}
Inman, D., \& Ali-Ha\"\i{}moud, Y. 2019, Phys. Rev. D, 100, 083528,
  \dodoi{10.1103/PhysRevD.100.083528}

\bibitem[{{Ioka} \& {M{\'e}sz{\'a}ros}(2005)}]{2005ApJ...619..684I}
{Ioka}, K., \& {M{\'e}sz{\'a}ros}, P. 2005, \apj, 619, 684,
  \dodoi{10.1086/426785}

\bibitem[{Ishiyama {et~al.}(2020)}]{Ishiyama:2020vao}
Ishiyama, T., {et~al.} 2020.
\newblock \doarXiv{2007.14720}

\bibitem[{Ivezic {et~al.}(2002)}]{Ivezic:2002gh}
Ivezic, Z., {et~al.} 2002, Astron. J., 124, 2364, \dodoi{10.1086/344069}

\bibitem[{{Kalaja} {et~al.}(2019){Kalaja}, {Bellomo}, {Bartolo}, {Bertacca},
  {Matarrese}, {Musco}, {Raccanelli}, \& {Verde}}]{2019JCAP...10..031K}
{Kalaja}, A., {Bellomo}, N., {Bartolo}, N., {et~al.} 2019, \jcap, 2019, 031,
  \dodoi{10.1088/1475-7516/2019/10/031}

\bibitem[{Kavanagh(2019)}]{bradley_j_kavanagh_2019_3538999}
Kavanagh, B.~J. 2019, bradkav/PBHbounds: Release version, 1.0,  Zenodo,
  \dodoi{10.5281/zenodo.3538999}

\bibitem[{{Kavanagh} {et~al.}(2018){Kavanagh}, {Gaggero}, \&
  {Bertone}}]{2018PhRvD..98b3536K}
{Kavanagh}, B.~J., {Gaggero}, D., \& {Bertone}, G. 2018, \prd, 98, 023536,
  \dodoi{10.1103/PhysRevD.98.023536}

\bibitem[{Kawasaki {et~al.}(2020)Kawasaki, Nakano, Nakatsuka, \&
  Sonomoto}]{Kawasaki:2020tbo}
Kawasaki, M., Nakano, W., Nakatsuka, H., \& Sonomoto, E. 2020.
\newblock \doarXiv{2010.13504}

\bibitem[{{Kuhlen} {et~al.}(2006){Kuhlen}, {Madau}, \&
  {Montgomery}}]{2006ApJ...637L...1K}
{Kuhlen}, M., {Madau}, P., \& {Montgomery}, R. 2006, \apjl, 637, L1,
  \dodoi{10.1086/500548}

\bibitem[{Leite {et~al.}(2017)Leite, Evoli, D'Angelo, Ciardi, Sigl, \&
  Ferrara}]{Leite:2017ekt}
Leite, N., Evoli, C., D'Angelo, M., {et~al.} 2017, Mon. Not. Roy. Astron. Soc.,
  469, 416, \dodoi{10.1093/mnras/stx805}

\bibitem[{Lidz \& Hui(2018)}]{Lidz:2018fqo}
Lidz, A., \& Hui, L. 2018, Phys. Rev., D98, 023011,
  \dodoi{10.1103/PhysRevD.98.023011}

\bibitem[{Maccio' {et~al.}(2008)Maccio', Dutton, \& Bosch}]{Maccio:2008pcd}
Maccio', A.~V., Dutton, A.~A., \& Bosch, F. C. v.~d. 2008, Mon. Not. Roy.
  Astron. Soc., 391, 1940, \dodoi{10.1111/j.1365-2966.2008.14029.x}

\bibitem[{{Mack} \& {Wyithe}(2012)}]{2012MNRAS.425.2988M}
{Mack}, K.~J., \& {Wyithe}, J. S.~B. 2012, \mnras, 425, 2988,
  \dodoi{10.1111/j.1365-2966.2012.21561.x}

\bibitem[{{Madau}(2018)}]{2018MNRAS.480L..43M}
{Madau}, P. 2018, MNRAS, 480, L43, \dodoi{10.1093/mnrasl/sly125}

\bibitem[{Madau \& Fragos(2017)}]{Madau:2016jbv}
Madau, P., \& Fragos, T. 2017, Astrophys. J., 840, 39,
  \dodoi{10.3847/1538-4357/aa6af9}

\bibitem[{{Makino} {et~al.}(1998){Makino}, {Sasaki}, \&
  {Suto}}]{1998ApJ...497..555M}
{Makino}, N., {Sasaki}, S., \& {Suto}, Y. 1998, \apj, 497, 555,
  \dodoi{10.1086/305507}

\bibitem[{{Meiksin}(2011)}]{2011MNRAS.417.1480M}
{Meiksin}, A. 2011, \mnras, 417, 1480, \dodoi{10.1111/j.1365-2966.2011.19362.x}

\bibitem[{Mena {et~al.}(2019)Mena, Palomares-Ruiz, Villanueva-Domingo, \&
  Witte}]{Mena:2019nhm}
Mena, O., Palomares-Ruiz, S., Villanueva-Domingo, P., \& Witte, S.~J. 2019,
  Phys. Rev., D100, 043540, \dodoi{10.1103/PhysRevD.100.043540}

\bibitem[{{Mesinger}(2019)}]{2019cosm.book.....M}
{Mesinger}, A. 2019, {The Cosmic 21-cm Revolution; Charting the first billion
  years of our universe} (AAS IOP Astronomy), \dodoi{10.1088/2514-3433/ab4a73}

\bibitem[{Mesinger {et~al.}(2011)Mesinger, Furlanetto, \&
  Cen}]{Mesinger:2010ne}
Mesinger, A., Furlanetto, S., \& Cen, R. 2011, Mon. Not. Roy. Astron. Soc.,
  411, 955, \dodoi{10.1111/j.1365-2966.2010.17731.x}

\bibitem[{{Monroy-Rodr{\'\i}guez} \& {Allen}(2014)}]{2014ApJ...790..159M}
{Monroy-Rodr{\'\i}guez}, M.~A., \& {Allen}, C. 2014, \apj, 790, 159,
  \dodoi{10.1088/0004-637X/790/2/159}

\bibitem[{{Murgia} {et~al.}(2019){Murgia}, {Scelfo}, {Viel}, \&
  {Raccanelli}}]{2019PhRvL.123g1102M}
{Murgia}, R., {Scelfo}, G., {Viel}, M., \& {Raccanelli}, A. 2019, \prl, 123,
  071102, \dodoi{10.1103/PhysRevLett.123.071102}

\bibitem[{Narayan \& Yi(1994)}]{Narayan:1994xi}
Narayan, R., \& Yi, I. 1994, Astrophys. J., 428, L13, \dodoi{10.1086/187381}

\bibitem[{Navarro {et~al.}(1997)Navarro, Frenk, \& White}]{Navarro:1996gj}
Navarro, J.~F., Frenk, C.~S., \& White, S. D.~M. 1997, Astrophys. J., 490, 493,
  \dodoi{10.1086/304888}

\bibitem[{Oguri {et~al.}(2018)Oguri, Diego, Kaiser, Kelly, \&
  Broadhurst}]{Oguri:2017ock}
Oguri, M., Diego, J.~M., Kaiser, N., Kelly, P.~L., \& Broadhurst, T. 2018,
  Phys. Rev. D, 97, 023518, \dodoi{10.1103/PhysRevD.97.023518}

\bibitem[{Park {et~al.}(2019)Park, Mesinger, Greig, \& Gillet}]{Park:2018ljd}
Park, J., Mesinger, A., Greig, B., \& Gillet, N. 2019, Mon. Not. Roy. Astron.
  Soc., 484, 933, \dodoi{10.1093/mnras/stz032}

\bibitem[{Peacock(1999)}]{Peacock:1999ye}
Peacock, J.~A. 1999, {Cosmological physics} (Cambridge University Press),
  \dodoi{10.1017/CBO9780511804533}

\bibitem[{Poulin {et~al.}(2017)Poulin, Serpico, Calore, Clesse, \&
  Kohri}]{Poulin:2017bwe}
Poulin, V., Serpico, P.~D., Calore, F., Clesse, S., \& Kohri, K. 2017, Phys.
  Rev., D96, 083524, \dodoi{10.1103/PhysRevD.96.083524}

\bibitem[{Press \& Schechter(1974)}]{Press:1973iz}
Press, W.~H., \& Schechter, P. 1974, Astrophys. J., 187, 425,
  \dodoi{10.1086/152650}

\bibitem[{Pritchard \& Loeb(2012)}]{Pritchard:2011xb}
Pritchard, J.~R., \& Loeb, A. 2012, Rept. Prog. Phys., 75, 086901,
  \dodoi{10.1088/0034-4885/75/8/086901}

\bibitem[{Ragagnin {et~al.}(2019)Ragagnin, Dolag, Moscardini, Biviano, \&
  D'Onofrio}]{Ragagnin:2018enf}
Ragagnin, A., Dolag, K., Moscardini, L., Biviano, A., \& D'Onofrio, M. 2019,
  Mon. Not. Roy. Astron. Soc., 486, 4001, \dodoi{10.1093/mnras/stz1103}

\bibitem[{Rees {et~al.}(1982)Rees, Phinney, Begelman, \&
  Blandford}]{Rees:1982pe}
Rees, M.~J., Phinney, E.~S., Begelman, M.~C., \& Blandford, R.~D. 1982, Nature,
  295, 17, \dodoi{10.1038/295017a0}

\bibitem[{{Sasaki} {et~al.}(2016){Sasaki}, {Suyama}, {Tanaka}, \&
  {Yokoyama}}]{2016PhRvL.117f1101S}
{Sasaki}, M., {Suyama}, T., {Tanaka}, T., \& {Yokoyama}, S. 2016, \prl, 117,
  061101, \dodoi{10.1103/PhysRevLett.117.061101}

\bibitem[{Semelin(2016)}]{Semelin:2015wca}
Semelin, B. 2016, Mon. Not. Roy. Astron. Soc., 455, 962,
  \dodoi{10.1093/mnras/stv2312}

\bibitem[{{Serpico} {et~al.}(2020){Serpico}, {Poulin}, {Inman}, \&
  {Kohri}}]{2020PhRvR...2b3204S}
{Serpico}, P.~D., {Poulin}, V., {Inman}, D., \& {Kohri}, K. 2020, Physical
  Review Research, 2, 023204, \dodoi{10.1103/PhysRevResearch.2.023204}

\bibitem[{Shakura \& Sunyaev(1973)}]{Shakura:1972te}
Shakura, N.~I., \& Sunyaev, R.~A. 1973, Astron. Astrophys., 24, 337

\bibitem[{Sheth {et~al.}(2001)Sheth, Mo, \& Tormen}]{Sheth:1999su}
Sheth, R.~K., Mo, H.~J., \& Tormen, G. 2001, Mon. Not. Roy. Astron. Soc., 323,
  1, \dodoi{10.1046/j.1365-8711.2001.04006.x}

\bibitem[{Sheth \& Tormen(1999)}]{Sheth:1999mn}
Sheth, R.~K., \& Tormen, G. 1999, Mon. Not. Roy. Astron. Soc., 308, 119,
  \dodoi{10.1046/j.1365-8711.1999.02692.x}

\bibitem[{{Shimabukuro} {et~al.}(2014){Shimabukuro}, {Ichiki}, {Inoue}, \&
  {Yokoyama}}]{2014PhRvD..90h3003S}
{Shimabukuro}, H., {Ichiki}, K., {Inoue}, S., \& {Yokoyama}, S. 2014, \prd, 90,
  083003, \dodoi{10.1103/PhysRevD.90.083003}

\bibitem[{{Shimabukuro} {et~al.}(2020{\natexlab{a}}){Shimabukuro}, {Ichiki}, \&
  {Kadota}}]{2020PhRvD.101d3516S}
{Shimabukuro}, H., {Ichiki}, K., \& {Kadota}, K. 2020{\natexlab{a}}, \prd, 101,
  043516, \dodoi{10.1103/PhysRevD.101.043516}

\bibitem[{{Shimabukuro} {et~al.}(2020{\natexlab{b}}){Shimabukuro}, {Ichiki}, \&
  {Kadota}}]{2020PhRvD.102b3522S}
---. 2020{\natexlab{b}}, \prd, 102, 023522, \dodoi{10.1103/PhysRevD.102.023522}

\bibitem[{Slatyer(2016)}]{Slatyer:2015kla}
Slatyer, T.~R. 2016, Phys. Rev., D93, 023521,
  \dodoi{10.1103/PhysRevD.93.023521}

\bibitem[{{Tanvir} {et~al.}(2009){Tanvir}, {Fox}, {Levan}, {Berger},
  {Wiersema}, {Fynbo}, {Cucchiara}, {Kr{\"u}hler}, {Gehrels}, {Bloom},
  {Greiner}, {Evans}, {Rol}, {Olivares}, {Hjorth}, {Jakobsson}, {Farihi},
  {Willingale}, {Starling}, {Cenko}, {Perley}, {Maund}, {Duke}, {Wijers},
  {Adamson}, {Allan}, {Bremer}, {Burrows}, {Castro-Tirado}, {Cavanagh}, {de
  Ugarte Postigo}, {Dopita}, {Fatkhullin}, {Fruchter}, {Foley}, {Gorosabel},
  {Kennea}, {Kerr}, {Klose}, {Krimm}, {Komarova}, {Kulkarni}, {Moskvitin},
  {Mundell}, {Naylor}, {Page}, {Penprase}, {Perri}, {Podsiadlowski}, {Roth},
  {Rutledge}, {Sakamoto}, {Schady}, {Schmidt}, {Soderberg}, {Sollerman},
  {Stephens}, {Stratta}, {Ukwatta}, {Watson}, {Westra}, {Wold}, \&
  {Wolf}}]{2009Natur.461.1254T}
{Tanvir}, N.~R., {Fox}, D.~B., {Levan}, A.~J., {et~al.} 2009, \nat, 461, 1254,
  \dodoi{10.1038/nature08459}

\bibitem[{{Tisserand} {et~al.}(2007){Tisserand}, {Le Guillou}, {Afonso},
  {Albert}, {Andersen}, {Ansari}, {Aubourg}, {Bareyre}, {Beaulieu}, {Charlot},
  {Coutures}, {Ferlet}, {Fouqu{\'e}}, {Glicenstein}, {Goldman}, {Gould},
  {Graff}, {Gros}, {Haissinski}, {Hamadache}, {de Kat}, {Lasserre}, {Lesquoy},
  {Loup}, {Magneville}, {Marquette}, {Maurice}, {Maury}, {Milsztajn}, {Moniez},
  {Palanque-Delabrouille}, {Perdereau}, {Rahal}, {Rich}, {Spiro},
  {Vidal-Madjar}, {Vigroux}, {Zylberajch}, \& {EROS-2
  Collaboration}}]{2007A&A...469..387T}
{Tisserand}, P., {Le Guillou}, L., {Afonso}, C., {et~al.} 2007, \aap, 469, 387,
  \dodoi{10.1051/0004-6361:20066017}

\bibitem[{{Toma} {et~al.}(2011){Toma}, {Sakamoto}, \&
  {M{\'e}sz{\'a}ros}}]{2011ApJ...731..127T}
{Toma}, K., {Sakamoto}, T., \& {M{\'e}sz{\'a}ros}, P. 2011, \apj, 731, 127,
  \dodoi{10.1088/0004-637X/731/2/127}

\bibitem[{{Vasiliev} \& {Shchekinov}(2012)}]{2012arXiv1205.1204V}
{Vasiliev}, E.~O., \& {Shchekinov}, Y.~A. 2012, arXiv e-prints,
  arXiv:1205.1204.
\newblock \doarXiv{1205.1204}

\bibitem[{Villanueva-Domingo(2021)}]{pablo_villanueva_domingo_2021_4707447}
Villanueva-Domingo, P. 2021, PabloVD/21cmForest\_PBH: 21cmForest\_PBH, v1.0.0,
  Zenodo, \dodoi{10.5281/zenodo.4707447}

\bibitem[{Villanueva-Domingo {et~al.}(2021)Villanueva-Domingo, Mena, \&
  Palomares-Ruiz}]{Villanueva-Domingo:2021spv}
Villanueva-Domingo, P., Mena, O., \& Palomares-Ruiz, S. 2021.
\newblock \doarXiv{2103.12087}

\bibitem[{Wise {et~al.}(2014)Wise, Demchenko, Halicek, Norman, Turk, Abel, \&
  Smith}]{Wise:2014vwa}
Wise, J.~H., Demchenko, V.~G., Halicek, M.~T., {et~al.} 2014, Mon. Not. Roy.
  Astron. Soc., 442, 2560, \dodoi{10.1093/mnras/stu979}

\bibitem[{Xie \& Yuan(2012)}]{Xie:2012rs}
Xie, F.-G., \& Yuan, F. 2012, Mon. Not. Roy. Astron. Soc., 427, 1580,
  \dodoi{10.1111/j.1365-2966.2012.22030.x}

\bibitem[{Xu {et~al.}(2009)Xu, Chen, Fan, Trac, \& Cen}]{Xu_2009}
Xu, Y., Chen, X., Fan, Z., Trac, H., \& Cen, R. 2009, The Astrophysical
  Journal, 704, 1396, \dodoi{10.1088/0004-637x/704/2/1396}

\bibitem[{{Xu} {et~al.}(2011){Xu}, {Ferrara}, \& {Chen}}]{2011MNRAS.410.2025X}
{Xu}, Y., {Ferrara}, A., \& {Chen}, X. 2011, \mnras, 410, 2025,
  \dodoi{10.1111/j.1365-2966.2010.17579.x}

\bibitem[{Yuan \& Narayan(2014)}]{Yuan:2014gma}
Yuan, F., \& Narayan, R. 2014, Ann. Rev. Astron. Astrophys., 52, 529,
  \dodoi{10.1146/annurev-astro-082812-141003}

\end{thebibliography}

\end{document}